\documentclass[journal,onecolumn,draftclsnofoot,]{IEEEtran}
\usepackage{setspace}
\usepackage[colorlinks=true]{hyperref}
 \hypersetup{
     colorlinks=true,
     linkcolor=blue,
     filecolor=blue,
     citecolor = Magenta,      
     urlcolor=cyan,
     }
\usepackage[dvipsnames]{xcolor}

\DeclareMathAlphabet{\dutchcal}{U}{dutchcal}{m}{n}
\SetMathAlphabet{\dutchcal}{bold}{U}{dutchcal}{b}{n}
\DeclareMathAlphabet{\dutchbcal} {U}{dutchcal}{b}{n}
\usepackage[dvipsnames]{xcolor}
\usepackage{url}
\usepackage{graphicx}
\usepackage{enumitem}
\usepackage{centernot}
\usepackage{epstopdf}
\usepackage{csquotes}
\usepackage{subfig}
\usepackage{mathtools}
\usepackage{amssymb, amsmath,amsthm, amsfonts}
\usepackage{ifthen}
\usepackage{tikz}
\usepackage{cite}
\usepackage{verbatim}
\usepackage{pifont}
\usepackage{bm}  
\usepackage{stackengine}
\usepackage{bbm}

\allowdisplaybreaks
\usepackage{accents}
\patchcmd{\subsubsection}{\itshape}{\bfseries}{}{}

\usepackage{xpatch}

\makeatletter
\let\save@mathaccent\mathaccent
\newcommand*\if@single[3]{%
  \setbox0\hbox{${\mathaccent"0362{#1}}^H$}%
  \setbox2\hbox{${\mathaccent"0362{\kern0pt#1}}^H$}%
  \ifdim\ht0=\ht2 #3\else #2\fi
  }
\newcommand*\rel@kern[1]{\kern#1\dimexpr\macc@kerna}
\newcommand*\widebar[1]{\@ifnextchar^{{\wide@bar{#1}{0}}}{\wide@bar{#1}{1}}}
\newcommand*\wide@bar[2]{\if@single{#1}{\wide@bar@{#1}{#2}{1}}{\wide@bar@{#1}{#2}{2}}}
\newcommand*\wide@bar@[3]{%
  \begingroup
  \def\mathaccent##1##2{%
    \let\mathaccent\save@mathaccent
    \if#32 \let\macc@nucleus\first@char \fi
    \setbox\z@\hbox{$\macc@style{\macc@nucleus}_{}$}%
    \setbox\tw@\hbox{$\macc@style{\macc@nucleus}{}_{}$}%
    \dimen@\wd\tw@
    \advance\dimen@-\wd\z@
    \divide\dimen@ 3
    \@tempdima\wd\tw@
    \advance\@tempdima-\scriptspace
    \divide\@tempdima 10
    \advance\dimen@-\@tempdima
    \ifdim\dimen@>\z@ \dimen@0pt\fi
    \rel@kern{0.6}\kern-\dimen@
    \if#31
      \overline{\rel@kern{-0.6}\kern\dimen@\macc@nucleus\rel@kern{0.4}\kern\dimen@}%
      \advance\dimen@0.4\dimexpr\macc@kerna
      \let\final@kern#2%
      \ifdim\dimen@<\z@ \let\final@kern1\fi
      \if\final@kern1 \kern-\dimen@\fi
    \else
      \overline{\rel@kern{-0.6}\kern\dimen@#1}%
    \fi
  }%
  \macc@depth\@ne
  \let\math@bgroup\@empty \let\math@egroup\macc@set@skewchar
  \mathsurround\z@ \frozen@everymath{\mathgroup\macc@group\relax}%
  \macc@set@skewchar\relax
  \let\mathaccentV\macc@nested@a
  \if#31
    \macc@nested@a\relax111{#1}%
  \else
    \def\gobble@till@marker##1\endmarker{}%
    \futurelet\first@char\gobble@till@marker#1\endmarker
    \ifcat\noexpand\first@char A\else
      \def\first@char{}%
    \fi
    \macc@nested@a\relax111{\first@char}%
  \fi
  \endgroup
}
\makeatother

\makeatletter
\AtBeginDocument{\xpatchcmd{\@thm}{\thm@headpunct{.}}{\thm@headpunct{}}{}{}}
\makeatother
\DeclareSymbolFont{bbold}{U}{bbold}{m}{n}
\DeclareSymbolFontAlphabet{\mathbbold}{bbold}

\newtheorem{theorem}{Theorem}
\newtheorem{cor}[theorem]{Corollary}
\newtheorem{lemma}[theorem]{Lemma}
\newtheorem{prop}[theorem]{Proposition}

\newtheorem{remark}[theorem]{Remark}

\newtheorem{example}[theorem]{Example}

\newcommand\Item[1][]{%
  \ifx\relax#1\relax  \item \else \item[#1] \fi
  \abovedisplayskip=0pt\abovedisplayshortskip=0pt~\vspace*{-\baselineskip}}

\newcommand{\norm}[1]{\left\Vert#1\right\Vert}
\newcommand{\abs}[1]{\left\vert#1\right\vert}

\newcommand{\bra}[1]{\langle#1|}
\newcommand{\ket}[1]{|#1\rangle}
\newcommand{\ketbra}[2]{|#1\rangle \langle #2|}

\newcommand{\innp}[2]{\langle #1,#2\rangle}

\newcommand{\cA}{\mathcal{A}}
\newcommand{\cB}{\mathcal{B}}
\newcommand{\cC}{\mathcal{C}}
\newcommand{\cD}{\mathcal{D}}

\newcommand{\cF}{\mathcal{F}}
\newcommand{\cG}{\mathcal{G}}

\newcommand{\cI}{\mathcal{I}}
\newcommand{\cJ}{\mathcal{J}}

\newcommand{\cL}{\mathcal{L}}
\newcommand{\cM}{\mathcal{M}}

\newcommand{\cP}{\mathcal{P}}
\newcommand{\cQ}{\mathcal{Q}}

\newcommand{\cS}{\mathcal{S}}
\newcommand{\cT}{\mathcal{T}}

\newcommand{\cW}{\mathcal{W}}
\newcommand{\cX}{\mathcal{X}}
\newcommand{\cY}{\mathcal{Y}}

\newcommand{\HH}{\mathbb{H}}

\newcommand{\NN}{\mathbb{N}}

\newcommand{\RR}{\mathbb{R}}

\newcommand{\tr}[1]{\operatorname{Tr}{\left[#1\right]}}
\newcommand{\ptr}[2]{\operatorname{Tr}_{#2}{\left[#1\right]}}

\newcommand{\ind}{\mathbbm 1}

\newcommand{\qrel}[2]{\mathsf{D}\left(#1\middle\|#2\right)}

\newcommand\blfootnote[1]{%
	\begingroup
	\renewcommand\thefootnote{}\footnote{#1}%
	\addtocounter{footnote}{-1}%
	\endgroup
}

\definecolor{cblue}{rgb}{0.16, 0.32, 0.75}

\def\h2{\tilde h}

\def\hm1{\hat h_{-1}}

\title{Distributed Quantum Hypothesis Testing under Zero-rate Communication Constraints}
\author{Sreejith Sreekumar, Christoph Hirche, Hao-Chung Cheng, and Mario Berta}

\begin{document}
\maketitle
\vspace{-1 cm}

\begin{abstract}
    The trade-offs between error probabilities in quantum hypothesis testing are by now well-understood in the centralized setting, but much less is known for distributed settings. Here, we study a distributed binary hypothesis testing problem to infer a bipartite quantum state shared between two remote parties, where one of these parties communicates to the tester at (asymptotic) zero-rate, while the other party communicates to the tester at zero-rate or higher. As our main contribution, we derive an efficiently computable single-letter formula for the Stein's exponent of this problem, when the state under the alternative is the product of their marginals. For proving the converse direction of our result, we utilize a novel technique based on  reverse hypercontractivity of a quantum markov semigroup combined with the pinching method. For the general case with vanishing type I error probability, we show that the Stein's exponent when (at least) one of the parties communicates classically at zero-rate is given by a multi-letter expression involving regularized measured relative entropy maximized over a sub-class of binary outcome separable measurements. When the state under the alternative commutes  with a product eigenbasis  of the marginal states under the null and has a larger support, we show that the exponent is characterized as a max-min optimization of regularized measured relative entropy over a sub-class of local binary outcome  projective measurements. While this expression becomes single-letter for the fully classical case, we further prove that this already does not happen in the same way    for classical-quantum states  in general. The converse proof of the max-min characterization relies on an extension of the classical blowing-up lemma to bipartite quantum states satisfying the aforementioned commutativity condition,  which could  be of independent interest.
\end{abstract}



\section{Introduction}
\blfootnote{S. Sreekumar (email: sreejith.sreekumar@centralesupelec.fr) is with the Laboratoire Des Signaux Et Systèmes (L2S), CNRS, Centralesupélec, Université Paris-Saclay. This work was primarily done while he was at RWTH Aachen University, Germany. M. Berta is with the Institute for Quantum Information at RWTH Aachen University, Germany. Christoph Hirche is with the  Institute for Information Processing (tnt/L3S) at  University of Hanover, Germany. Hao-Chung Cheng is with the Department of Electrical Engineering, National Taiwan University, 
National Center for Theoretical Sciences, and 
Hon Hai (Foxconn) Quantum Computing Center, Taiwan.}

Quantum hypothesis testing to discriminate between two quantum states is perhaps the most  fundamental and simplest form of inference on a quantum system. 
The performance of quantum hypothesis testing is well-known and  characterized  by the Helstrom-Holevo test which achieves the  optimal trade-off between the type I and II error probabilities. Moreover, given an arbitrary number of independent and identical (i.i.d.) copies of quantum states, the best asymptotic rate of decay of error probabilities (or error-exponents) in different regimes of interest have also been ascertained \cite{Hiai-Petz-1991,Nussbaum-Szkola-2006,Audenaert-2008}. These error-exponents  have simple characterizations in terms of single-letter expressions involving quantum relative entropy \cite{Umegaki-62}  and its R\'{e}nyi generalizations \cite{Renyi-60,Petz1985Quasi-entropiesAlgebra,Petz1986Quasi-entropiesSystems,muller2013quantum,wilde2014strong}. However, such  expressions depend on the assumption that the tester, who performs the test, has direct access to the i.i.d. copies and can perform the optimal measurement to deduce the true hypothesis. We refer to such a scenario as the  \textit{centralized} setting.  
The situation dramatically changes when the tester 
has only remote or partial access to the  quantum (sub-)systems. 
In contrast to the centralized setting, the performance of quantum hypothesis testing   in such scenarios is much less understood.
\begin{figure}[t]
\centering
\includegraphics[trim=3.8cm 6cm 1cm 5cm, clip, width= 1.06\textwidth]{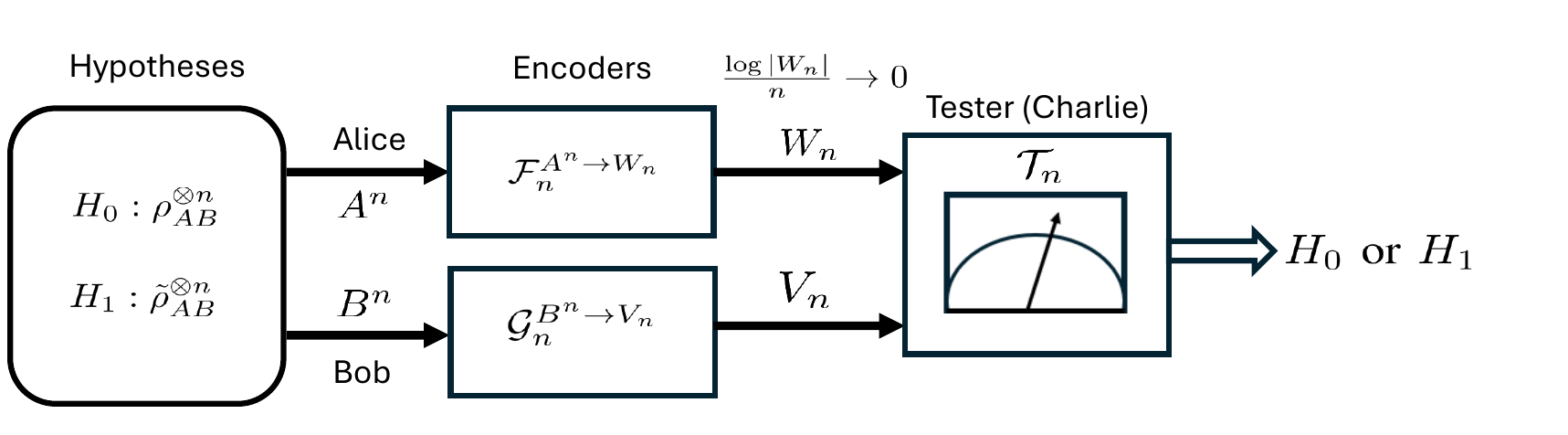}
\caption{Distributed quantum hypothesis testing under a zero-rate noiseless communication constraint. At least, one of the parties (Alice here) communicates at zero-rate to the tester (Charlie). The other party (Bob here)  communicates to Charlie at zero-rate or higher.} \label{Fig:DHT-zerorate}
\end{figure}

Here, we consider a distributed binary hypothesis testing problem  to discriminate between a bipartite quantum state shared between two parties, Alice and Bob,  as shown in Figure \ref{Fig:DHT-zerorate}. Each party has access to as many i.i.d. copies of the true state, $\rho_{AB}$ under the null or $\tilde \rho_{AB}$ under the alternative, as desired and can perform local quantum operations (e.g., local measurements)  on its own subsystem. The outputs of the local operation  are communicated to a tester, referred to as Charlie, under a constraint on the rate of communication. We focus primarily on the (asymptotically) zero-rate\footnote{Zero-rate means that the rate of communication vanishes asymptotically with respect to the number of copies of the state.} noiseless  communication  scenario, where at least one of the parties  communicates (classical or quantum information) to  Charlie at zero-rate and the other party is allowed to communicate (classical or quantum information) at zero-rate or higher. Zero-rate communication is  practically relevant in scenarios where communication is expensive, e.g.,  sensor networks or communication in an adversarial environment. As our main contribution, we characterize the \textit{Stein's} exponent, which specifies the optimal asymptotic rate of decay of the type II error probability for a fixed constraint on the type I error probability.   Our results generalize the corresponding expressions derived in \cite{Han-1987} and \cite{Shalaby-pap} for the classical case to the quantum setting.

  In contrast to the centralized scenario, design of an optimal  hypothesis testing scheme in distributed settings also involves an optimization over all feasible encoders (in addition to the test statistic). In distributed quantum settings, there is an additional complexity  arising due to constraints on the kind of operations that could be performed on the system. For instance,  such constraints could occur naturally due to restriction on the set of feasible measurements 
  (e.g., due to geographic separation) or the distributed processing of information post-measurement. This makes the characterization of Stein's exponent more challenging, and even classically,  single-letter expressions are known only in special cases. One such important setting is the zero-rate regime, which partly  motivates our study.

We show that the Stein's exponent under a type I error probability constraint $\epsilon$, denoted by $\theta(\epsilon, \rho_{AB}, \tilde \rho_{AB})$,   is given by a single-letter formula when the state under the alternative is the product of its marginals ($\tilde \rho_{AB}=\tilde\rho_{A} \otimes \tilde\rho_{B}$):
   \begin{align}
 \theta(\epsilon, \rho_{AB}, \tilde \rho_{A} \otimes \tilde \rho_{B})   &=\qrel{\rho_{A}}{\tilde \rho_A}+\qrel{\rho_{B}}{\tilde \rho_B}, ~\forall~ \epsilon \in (0,1). \label{eq:singletint}
   \end{align}  
The proof of \eqref{eq:singletint}  as usual involves establishing achievability (lower bound) and a converse (upper bound). The achievability relies on  arguments based on quantum-typicality similar to those in \cite{Bjela-Schultze-2012}. For establishing the converse bound, we leverage a novel proof technique based on reverse hypercontractivity of a quantum markov semigroup (QMS) \cite{Beigi-2020,Cheng-2021} in conjunction with the pinching technique of \cite{Hayashi_2002}. Notably, the proof technique leads to a second-order converse (see Lemma \ref{Lem:strong-converse-revhyp}) that goes beyond the setting of classical-quantum (CQ) states, for which a second-order converse  in the positive rate setting was derived in \cite{Cheng-2021}. 
  
 For the general case, $\theta(\epsilon, \rho_{AB}, \tilde \rho_{AB})$  admits a multi-letter characterization (see \eqref{eq:multilett-char}) in the regime of vanishing type I error probability constraint, which can be shown via standard techniques and is hard to compute in general. However, this characterization can be simplified in certain instances.  Of particular interest is the scenario where at least one of the parties communicates classically at zero-rate to Charlie, in which case we denote  $ \theta(\epsilon,\rho_{AB}, \tilde \rho_{AB})$ by $ \theta_{\mathsf{C}}(\epsilon,\rho_{AB}, \tilde \rho_{AB})$. 
For this, we characterize $ \lim_{\epsilon \downarrow 0^+}\theta_{\mathsf{C}}(\epsilon, \rho_{AB}, \tilde \rho_{AB})$ in terms of a multi-letter  expression (see \eqref{eq:limitexp-sepzrexp}) involving regularized measured relative entropy maximized over a class of binary separable measurements. Additionally, when $\rho_{AB}$ and $\tilde \rho_{AB}$ are such that a product eigenbasis of $\rho_A \otimes \rho_B$ commutes with $\tilde \rho_{AB}$ and the support condition $\rho_{A} \otimes \rho_{B} \ll \tilde \rho_{AB}$ holds, then the exponent is given by a multi-letter expression  (see \eqref{eq:limitexplocstein}) involving max-min optimization of  regularized measured relative entropy. Here, the maximization is over the class of null marginal compatible local measurements and the minimization is over a set of auxiliary quantum states whose marginals coincide with those of the tensor product state under the null hypothesis. 

For proving achievability in \eqref{eq:limitexplocstein}, we consider a \textit{local scheme} where each encoder applies a sequence of identical local (composite) measurements acting on multiple i.i.d. copies of the quantum state and communicates the measurement outcomes via an optimal zero-rate classical scheme. In fact, the optimal exponent is achieved by a single bit of classical communication each from Alice and Bob to Charlie. 
For establishing the upper bound (converse) in terms of \eqref{eq:limitexplocstein}, we derive a non-asymptotic strong converse\footnote{Strong converse here refers to the optimal  Stein's exponent being independent of the constraint, $\epsilon \in (0,1)$, on the type I error probability.} via a generalization of the classical blowing-up lemma \cite{Ahlswede-Gacs-Korner-1976,Csiszar-Korner,Marton-1986}. Specifically, this lemma holds under the support condition $\rho_{A} \otimes \rho_{B} \ll \tilde \rho_{AB}$ together with the commutativity condition that a product eigenbasis of $\rho_A \otimes \rho_B$ commutes with $\tilde \rho_{AB}$ (see Theorem \ref{Thm:Steinsexp-classical-zerorate}$(i)$).  The lower bound (achievability) in terms of \eqref{eq:limitexplocstein} can be shown using the local scheme as before, and matches asymptotically with the upper bound, thus characterizing the Stein's exponent in terms of the desired  expression.  
We also show via an example of CQ states (see Proposition \ref{Prop:counterexcqstate}) that $\theta_{\mathsf{C}}(\epsilon, \rho_{AB}, \tilde \rho_{AB})$ does not coincide, in general, with the natural quantum analogue of the single-letter expression characterizing it when $ \rho_{AB}$ and $ \tilde \rho_{AB}$ are classical states.  We further  highlight via an example that the Stein's exponent with single qudit quantum communication each from Alice and Bob to Charlie can be infinitely better than $\theta_{\mathsf{C}}(\epsilon, \rho_{AB}, \tilde \rho_{AB})$ (see Example \ref{ex:zeroratecvsq} below). This exhibits an interesting dichotomy of sorts between the classical and quantum realm. 

\medskip

\noindent
\textbf{NOTE:} Unfortunately, there was an error in the proof of the quantum version of the blowing-up lemma given in the published versions of the paper \cite{DQHT-2025-isit,DQHT-2025-AHP}, i.e., the penultimate inequality in \cite[Proof of Lemma 4]{DQHT-2025-AHP} and similar inequality  in the \cite[Proof of Lemma 16]{DQHT-2025-AHP}. We are grateful to Alptug Aytekin and  Sennur Ulukus of the University of Maryland, who pointed out the error to us recently via private communication. To resolve this error, the corrected lemma assumes that $\tilde\rho_{AB}$ commutes with the rank-one product projectors of suitable local eigenbasis of $\rho_A$ and $\rho_B$ on their supports. This product-eigenbasis commutativity condition appears in Theorem~\ref{Thm:Steinsexp-classical-zerorate}(i) and in results whose proof uses this lemma. Notably, we are able to retain without the commutativity assumption our prior result on single-letterization of Stein's exponent when the state under the alternative is a product of its marginals (Theorem \ref{Thm:Steinsexp-zerorate}). We do this by presenting an alternative novel proof of the converse direction based on a  technique combining reverse hypercontractivity of QMSs and quantum pinching.

\subsection{Related Work}\label{Sec:Relwork}
While the literature on centralized hypothesis testing is extensive in both statistics as well as classical and quantum information theory, here we focus only on the distributed setting. Distributed hypothesis testing with the goal of testing for the joint distribution of data  has been an active topic of research in classical information theory and statistics,  with the characterization of Stein's exponent and other performance metrics explored in various multi-terminal settings. A multi-letter characterization  of the Stein's exponent  in a two-terminal setting where the communication happens over a noiseless (positive) rate-limited channel was first established in \cite{Ahlswede-Csiszar}. Invoking the blowing-up lemma, a strong converse was also proved  under a positivity assumption on the probability distribution under the alternative. Moreover, it was shown that the exponent has a single-letter formula for testing against independence,  in which the joint distribution under the alternative is the product of its marginals. Further, a single-letter lower bound on the exponent for the general case was derived in \cite{Ahlswede-Csiszar,Han-1987} using a scheme based on data compression  which was subsequently improved in \cite{Shimokawa} by using binning to reduce the rate of communication (see \cite{Watanabe-2022,Kochman-Wang-2023} which shows the suboptimality of this scheme). In the other direction, improved single-letter upper bounds on the exponent have been established  in \cite{Rahman-Wagner,Hadar-Liu-Polyansky-2019,Kochman-24}. Extensions of the basic problem to a successive refinement model\cite{Tian-Chen-2008}, interactive settings \cite{Xiang-Kim-2,Katz-collab}, noisy communication channels \cite{Sree_isit17,SD_ISIT2020,SD_2020,Sadaf-Wigger-HTN}, and networks involving more than two terminals \cite{Rahman-Wagner,Zhao-Lai, Wigger-Timo,  Zhao-18, Sadaf-Wigger-Li, Escamilla-2020, Yunus-22,Zaidi-23,Hamad-23} have also been explored.

Distributed hypothesis testing under a zero-rate noiseless communication constraint  was introduced in \cite{Han-1987}, where a one-bit communication scheme indicating whether the observed sequence at each encoder is typical or not was proposed.  Subsequently, the optimality of the one-bit scheme along with a strong converse was established in \cite{Shalaby-pap} by leveraging the blowing-up lemma.  The trade-off between the type I and type II error-exponents in the same setting under a positive and zero-rate communication constraint was first explored in \cite{HK-1989}, where inner bounds were established (see also \cite{Han-Amari-1998}, and \cite{Zhao-Lai-2015,Haim-Kochman-16,Watanabe-18,WKJ-2017,WKW_isit19,Xu-2022,SD-2023} for more recent progress and extensions). Distributed hypothesis testing under additional constraints such as privacy,  security, or lossy data compression have also been investigated \cite{Katz-estdetjourn,Maggie-Pablo,SS-2018-privacy-conf,GSSV-2018, SCG-2019, SD-10-security}.

In contrast, analogous problems in quantum settings have been much less explored. The only exception that we are aware of is \cite{Cheng-2021}, which among other contributions, proved a strong converse establishing a single-letter characterization  of the Stein's exponent  for testing against independence. Here, we consider the first quantum analogue of the problem studied in  \cite{Han-1987,Shalaby-pap}.   We briefly compare our work with \cite{Cheng-2021}. While \cite{Cheng-2021} treats   testing against independence under a positive rate communication constraint, we work in the zero-rate setting allowing arbitrary marginals $\tilde \rho_A$ and $\tilde \rho_B$. Also, our proofs are markedly different from those in \cite{Cheng-2021}.  Notably, the strong converse for Stein's exponent characterization in  \eqref{eq:singletint} does not require  $\rho_{AB}$ to be a classical-quantum state $\rho_{XB}$ as required in \cite{Cheng-2021}.

The remainder of the paper is organized as follows. We formulate the problem of quantum distributed hypothesis testing under zero-rate communication constraints in Section \ref{Sec:Probform}. The main results are presented in Section \ref{Sec:mainres} and its proofs are given in Section \ref{Sec:Proofs}.  Concluding remarks along with some avenues for further research are provided  in Section \ref{Sec:concrem}.   
\subsection{Notation}
 The set of (linear) operators from a finite dimensional complex Hilbert space  $\HH_d$ of dimension $d$ to itself is denoted by  $\cL(\HH_d)$. 
$\tr{\cdot}$, $\innp{\cdot}{\cdot}$, and $\norm{\cdot}_p$ for $p \geq 1$ signifies the trace operation, inner product,  and Schatten $p$-norm, respectively. The
set of density operators (or quantum states) on $\HH_d$, i.e., positive semi-definite operators with unit trace is denoted by $\cS_d $.     The notation $\leq$ denotes the L\"{o}wner partial order in the context of operators, i.e., for Hermitian operators $H_1,H_2$, $H_1 \leq H_2$ means that $H_2-H_1$ is positive semi-definite. $\ind_{\cX}$ denotes the indicator of a set $\cX$ and $I$ denotes the identity operator. $H_1 \ll H_2$ designates that the support of $H_1$ is contained in that of $H_2$. 
For reals $a,b$, $a \wedge b :=\min\{a,b\}$, $a \vee b :=\max\{a,b\}$,  and for integers $m,n $ with $m \leq n$,  $[m:n] :=\{m,m+1,\ldots,n\}$. For a finite discrete set $\cX$, a positive operator-valued measure (POVM)  indexed by $\cX$ is a set $\cM=\{M_x\}_{x \in \cX}$ such that $ M_x  \geq 0$ for all $x$ and $\sum_{x \in \cX} M_x =I$. Such a  POVM  induces a measurement channel (quantum to classical channel) specified by
\begin{align}
&\cM(\omega):=\sum_{x \in \cX} \tr{M_x \omega} \ket{x}\bra{x}, ~~~ \omega \in \cL(\HH_{d}).\notag
\end{align}
For composite systems, we indicate the labels of the sub-systems involved as subscript or superscript wherever convenient, e.g., $\rho_{AB}$ for the  state of bipartite system $AB$.
$\mathsf{LO}_n(A^n:B^n)$ denotes the set of local POVM's on the bipartite system $A^nB^n$. This is abbreviated as $\mathsf{LO}_n$ when the systems are evident from context,  and further as $\mathsf{LO}$ when $n=1$. The marginals  of a bipartite linear operator $\omega_{AB}$ are defined as usual via partial trace operation,  e.g., $\omega_A:=\ptr{\omega_{AB}}{B}$, where $\ptr{\omega_{AB}}{B}$ denotes the partial trace of $\omega_{AB}$ with respect to system $B$.

\section{Problem Formulation }\label{Sec:Probform}
We consider a distributed binary hypothesis testing problem as shown in Fig. \ref{Fig:DHT-zerorate}, where Alice and Bob share $n$ identical copies of a bipartite quantum system $AB$ such that $A$ is with Alice and $B$ is with Bob, respectively. Assume that $A$ and $B$ are finite dimensional with size $d_A$ and $d_B$, respectively. The goal is to ascertain the true joint state on $AB$ at Charlie,  known to be either $\rho_{AB}$ or $\tilde {\rho}_{AB}$. This  may be formulated as a binary hypothesis testing problem with the following null and alternative hypotheses:
\begin{subequations}\label{eq:hyptestquant}
 \begin{align}
   &  \mathsf{H}_0 : \mbox{State on } A^nB^n \mbox{ is } \rho_{AB}^{\otimes n}, \\
   &  \mathsf{H}_1:  \mbox{State on } A^nB^n \mbox{ is } \tilde{\rho}_{AB}^{\otimes n}.
 \end{align}      
 \end{subequations}
 To perform this task, Alice and Bob are allowed to perform local operations on $A^n$ and $B^n$ and communicate with Charlie.  
 Alice and Bob perform encoding by applying completely positive trace-preserving (CPTP) linear maps (or quantum channels) $\cF_n=\cF_n^{A^n \to W_n}$ and $\cG_n=\cG_n^{B^n \to V_n}$, respectively, with $|W_n| \wedge |V_n|>1$. Alice  sends $W_n$ and Bob sends $V_n$ to Charlie over their respective noiseless channels. 
Let $\sigma_{W_n V_n}$  denote the state of $(W_n,V_n)$ under the null and $\tilde \sigma_{W_n V_n}$ denote the state of the same systems under the alternative.  Charlie  applies a binary outcome POVM  $\cT_n=\{T_n,I-T_n\}$ on $(W_n,V_n)$ to decide which hypothesis is true. The type I and type II error probabilities achieved by the test $\cT_n$ are 
\begin{align}
    \alpha_n(\cF_n,\cG_n,\cT_n)&=\tr{(I-T_n) \sigma_{W_n V_n}}, \notag \\
     \mbox{and  }\beta_n(\cF_n,\cG_n,\cT_n)&=\tr{T_n \tilde \sigma_{W_n V_n}}, \notag
\end{align} 
respectively. 
We assume that at least one of the parties communicates at zero-rate to Charlie, which we take to be Alice without loss of generality.  The  zero-rate communication constraint at Alice means that 
    \begin{align}
   \lim_{n \rightarrow \infty} \frac{1}{n}\log |W_n| = 0, \label{eq:zeroratequant}
    \end{align}
     where $|W_n|$ denotes the dimension of system $W_n$. 
We are interested in   characterizing the Stein's exponent defined as 
\begin{align}
   \theta(\epsilon,\rho_{AB}, \tilde \rho_{AB})&:= \lim_{n \rightarrow \infty} -\frac{\log \bar{\beta}_n(\epsilon)}{n}, \label{eq:Steinsexp} 
\end{align}    
if the above limit exists, where 
\begin{align}
\bar{\beta}_n(\epsilon)&:=\inf_{\cF_n,\cG_n,\cT_n}\{\beta_n(\cF_n,\cG_n,\cT_n): \alpha_n(\cF_n,\cG_n,\cT_n) \leq \epsilon\}. \notag
\end{align}
When Alice 
(or Bob or both) communicates classically at zero-rate to Charlie, we denote  $ \theta(\epsilon,\rho_{AB}, \tilde \rho_{AB})$ by $ \theta_{\mathsf{C}}(\epsilon,\rho_{AB}, \tilde \rho_{AB})$.  In this case, Alice's encoder $\cF_n$  is specified by a measurement channel $ \cM_n^{A^n \to X_n}(\cdot)$ with $W_n=X_n$ such that 
\begin{align}
    \cM_n^{A^n \to X_n}(\omega^{A^n}_n)= \sum_{x_n \in \cX_n}\tr{M_{x_n} \omega^{A^n}_n}\ket{x_n}\bra{x_n}, \label{eq:povmop}
\end{align}
where  $\cM_n=\{M_{x_n}\}_{x_n \in \cX_n}$ denotes a POVM with $|\cX_n|>1$ and
    \begin{align}
   \lim_{n \rightarrow \infty} \frac{1}{n}\log |\cX_n|= 0. \notag
    \end{align}
 If Bob communicates classically, we will use the variable $Y_n$ in place of $V_n$. Here,  $Y_n$ is obtained as the output of the measurement channel $ \cM_n^{B^n \to Y_n}(\cdot)$ similar to \eqref{eq:povmop}.   
When $B^n$ is directly accessible to Charlie, e.g., via teleportation  or quantum communication at a sufficiently large rate, then Bob may be identified  with Charlie. In particular, an upper bound on the Stein's exponent in this scenario will be an upper bound for the case when Bob communicates with Charlie.

\subsection{Existing Results}
One may obtain a multi-letter characterization of $\theta(\epsilon, \rho_{AB}, \tilde \rho_{AB})$ in the limit $\epsilon \downarrow 0^+$ by an adaptation of the proof of  \cite[Theorem 1]{Ahlswede-Csiszar} to the quantum setting. Following similar steps as in \cite{Cheng-2021}  yields the multi-letter characterization
\begin{align}
\lim_{\epsilon \downarrow 0^+}\theta(\epsilon, \rho_{AB}, \tilde \rho_{AB})= \lim_{n \rightarrow \infty} \sup_{\cF_n,\cG_n}\frac{1}{n} \qrel{\sigma_{W_nV_n}}{\tilde \sigma_{W_n V_n}}, \label{eq:multilett-char}
\end{align}
where the optimization is over all  encoders $\cF_n$ and $\cG_n$ as specified above. Since the set of all such encoders grows exponentially in $n$, the optimization in the right hand side (RHS)  of \eqref{eq:multilett-char} quickly becomes intractable as $n$ increases.

In the following, we will obtain a  relatively simpler characterization of $\theta(\epsilon, \rho_{AB}, \tilde \rho_{AB})$. 
Before that, it is instructive to  discuss the classical case. Therein, the goal is to test for the joint distribution of bipartite data $(X,Y)$. The relevant hypothesis test is given by
\begin{align} 
   & H_0: (X^n,Y^n) \sim p_{XY}^{\otimes n}, \notag \\
   & H_1: (X^n,Y^n) \sim \tilde p_{XY}^{\otimes n}, \notag
\end{align}
where  $p_{XY}$ and $\tilde p_{XY}$ denotes the probability mass function of the data under the null and alternative, respectively. 
An optimal scheme (see \cite{Han-1987}) achieving the Stein's exponent for this hypothesis test is as follows: Each party  locally tests whether its  observed sequence is typical according to the marginals under the null-hypothesis, i.e., Alice checks for $p_X$-typicality and Bob checks for  $p_Y$-typicality (see, e.g., \cite{Csiszar-Korner}, for the definition of typicality). Both parties then send a one bit message each to the tester informing whether local typicality test is successful. Upon receiving these messages, the tester  declares $H_0$ if both the local tests are successful. By standard typicality arguments in information theory, the type I error-probability  tends to zero asymptotically. Moreover, an application of Sanov's theorem 
 shows that this scheme asymptotically achieves the type II error-exponent  
\begin{align}
 \theta_{\mathsf{ZRC}}^{\star}(p_{XY}, \tilde p_{XY}):=   \min_{\hat p_{XY}: \hat p_X=p_X, \hat p_Y=p_Y } \mathsf{D}\big(\hat p_{XY}\|\tilde p_{XY}\big). \label{eq:classicaltestscheme}
\end{align}
In \cite{Shalaby-pap}, it was subsequently shown that $\theta(\epsilon, p_{XY}, \tilde p_{XY}) =\theta_{\mathsf{ZRC}}^{\star}(p_{XY}, \tilde p_{XY})$  for all $0 < \epsilon<1$ when $\tilde p_{XY}>0$. Note that computing $\theta_{\mathsf{ZRC}}^{\star}(p_{XY}, \tilde p_{XY})$ involves minimizing a convex function $\mathsf{D}\big(\hat p_{XY}\|\tilde p_{XY}\big)$ of $\hat p_{XY}$ over the convex set $ \{\hat p_{XY}: \hat p_X=p_X, \hat p_Y=p_Y\}$. This  is a convex optimization problem with linear constraints and number of variables equal to $|\cX||\cY|-1$. In particular, the number of variables do not grow with $n$, i.e., $\theta_{\mathsf{ZRC}}^{\star}(p_{XY}, \tilde p_{XY})$ is a single-letter expression, which  can be  computed efficiently.    In the next section, we will obtain a generalization of the aforementioned result for some specific instances in the quantum setting.

\section{Main Results}\label{Sec:mainres}
The following theorem which characterizes the Stein's exponent $\theta(\epsilon, \rho_{AB}, \tilde \rho_{A} \otimes \tilde \rho_{B})$ is one of our main results.  
\begin{theorem}[Single-letter zero-rate Stein's exponent]\label{Thm:Steinsexp-zerorate}
 Let $\tilde \rho_{AB}=\tilde \rho_{A}  \otimes \tilde \rho_{B}$. 
Then, for  all $\epsilon \in (0,1)$ and any $\rho_{AB}$,   
   \begin{align}
 \theta(\epsilon, \rho_{AB}, \tilde \rho_{AB}) =\qrel{\rho_{A}}{\tilde \rho_A}+\qrel{\rho_{B}}{\tilde \rho_B}. \label{eq: steinexp-productalt}
   \end{align}
\end{theorem} 
The proof of Theorem \ref{Thm:Steinsexp-zerorate} is given in Section \ref{Sec:Thm:Steinsexp-zerorate-proof}. Achievability of the single-letter expression in the RHS of \eqref{eq: steinexp-productalt} relies on quantum-typicality based arguments similar to those in \cite{Bjela-Schultze-2012}. To prove the converse direction, we derive a non-asymptotic second order converse (see Lemma \ref{Lem:strong-converse-revhyp} below) by using a novel pinching technique coupled with the quantum reverse hypercontractivity result for tensor products of the generalized depolarizing semigroup \cite{Beigi-2020}.

Note that Theorem \ref{Thm:Steinsexp-zerorate}  implies that for  \textit{testing against  independence}, $\theta(\epsilon, \rho_{AB},  \rho_{A} \otimes \rho_{B})  
 =0$. To compare this with the performance in the centralized setting, recall that the Stein's exponent in the latter case is equal to the mutual information evaluated with respect to  $\rho_{AB}$ (see, e.g., \cite{Hayashi-Tomamichel-16,Berta-2021} for a composite hypothesis testing version), which is positive except in the trivial case  $\rho_{AB}=\rho_A \otimes \rho_B$.   Hence, there is a strict gap between the performance under zero-rate  and centralized setting as expected.

Next, we obtain multi-letter characterizations of $\theta_{\mathsf{C}}(\epsilon, \rho_{AB}, \tilde \rho_{AB})$.   For this, we need to introduce some further notations.   
Let  $ \mathsf{PLO}_n$ be the set of local projective-valued measures (PVMs)  of the form  $\cM_n=\cP_n^{A^n} \otimes \cP_n^{B^n}$,  where $\cP_n^{A^n}=\{P_{x}^{A^n}=\ket{x}\bra{x}\}_{x \in [1:d_A^n]}$  and $\cP_n^{B^n}=\{P_{y}^{B^n}=\ket{y}\bra{y}\}_{y \in [1:d_B^n]}$  are  sets of orthogonal rank-one projections on $\HH_{d_A^n}$ and $\HH_{d_B^n}$, respectively.  Define the marginal-compatible subclass
\begin{align}
\mathsf{PLO}^{\rho}_n:=\Big\{\cP_n^{A^n}\otimes\cP_n^{B^n}\in\mathsf{PLO}_n:\ [P_x^{A^n},\rho_A^{\otimes n}]=0,\ [P_y^{B^n},\rho_B^{\otimes n}]=0\ \text{for all }x,y\Big\}.
\label{eq:rho-compatible-PLO}
\end{align}
and set
\begin{subequations} \label{eq:optimalmultlettexp}
 \begin{align}
\theta^{\star}_{\mathsf{PLO}\text{-}{\rho}}(\rho_{AB},\tilde\rho_{AB})
&:=\lim_{n\to\infty}\sup_{\cM_n\in\mathsf{PLO}^{\rho}_n}\inf_{\hat\rho^n_{AB}\in\cD_n(\rho_{AB})}
\frac{\qrel{\cM_n(\hat\rho^n_{AB})}{\cM_n(\tilde\rho_{AB}^{\otimes n})}}{n}\label{eq:limitexplocstein} \\
   &= \sup_{n  \in \NN}\sup_{\cM_n \in \mathsf{PLO}^{\rho}_n} \inf_{\hat \rho^n_{AB} \in \cD_n(\rho_{AB})} \frac{\qrel{\cM_n(\hat \rho^n_{AB})}{\cM_n(\tilde \rho_{AB}^{\otimes n})}}{n}, \label{eq:supadditexp-locstein} 
   \end{align} 
   \text{where}
\begin{align}
\cD_n(\rho_{AB})&:= \big\{\hat \rho^n_{AB}: \hat \rho_{A}^n=\rho_A^{\otimes n}, \hat \rho_{B}^n=\rho_B^{\otimes n}\big\}. \label{eq:minimizset}
   \end{align} 
\end{subequations}
Also, let 
\begin{align}
\theta^{\star}_{\mathsf{BSEP-ZR}}(\rho_{AB},\tilde \rho_{AB}) &:= \limsup_{n \rightarrow \infty} \sup_{\cM_n \in  \mathsf{BSEP-ZR_n}} \frac{\qrel{\cM_n( \rho_{AB}^{\otimes n})}{\cM_n(\tilde \rho_{AB}^{\otimes n})}}{n}, \label{eq:limitexp-sepzrexp}
\end{align}
where  
$\mathsf{BSEP-ZR_n}$ is the set of measurement channels induced by binary outcome separable POVMs of the form
\begin{align}
   \left\{ \sum_{x \in \cX_n} M_x^{A^n} \otimes M_x^{B^n}, I_{A^nB^n}-\sum_{x \in \cX_n} M_x^{A^n} \otimes M_x^{B^n} \right\}, \notag
\end{align}
for some set $\cX_n$ satisfying $\lim_{n \rightarrow \infty} \log |\cX_n|/n =0$. Note that $\theta^{\star}_{\mathsf{BSEP-ZR}}(\rho_{AB},\tilde \rho_{AB})$ does not have the inner minimization with respect to $\hat \rho^n_{AB} \in \cD_n(\rho_{AB})$ in contrast to the definition of 
$\theta^{\star}_{\mathsf{PLO}\text{-}{\rho}}$. 
   \begin{remark}[Super-additivity of max-min measured relative entropy] \label{eq:limexiststeinexp}
The existence of the limit in \eqref{eq:limitexplocstein} and the final equality in \eqref{eq:supadditexp-locstein} is a consequence of the following super-additivity property and Fekete's lemma:
\begin{align}
&\sup_{\cM_{n+k}\in\mathsf{PLO}_{n+k}^{\rho}}
 \inf_{\hat\rho^{\,n+k}_{AB}\in\cD_{n+k}(\rho_{AB})}
 \qrel{\cM_{n+k}(\hat\rho^{\,n+k}_{AB})}
       {\cM_{n+k}(\tilde\rho_{AB}^{\otimes(n+k)})}
\notag\\
&\qquad\ge
\sup_{\cM_n\in\mathsf{PLO}_{n}^{\rho}}
 \inf_{\hat\rho^{\,n}_{AB}\in\cD_n(\rho_{AB})}
 \qrel{\cM_n(\hat\rho^{\,n}_{AB})}
       {\cM_n(\tilde\rho_{AB}^{\otimes n})}
+
\sup_{\cM_k\in\mathsf{PLO}_{k}^{\rho}}
 \inf_{\hat\rho^{\,k}_{AB}\in\cD_k(\rho_{AB})}
 \qrel{\cM_k(\hat\rho^{\,k}_{AB})}
       {\cM_k(\tilde\rho_{AB}^{\otimes k})}.
\notag
\end{align}
Indeed, fix $\cM_n\in\mathsf{PLO}^{\rho}_n$ and $\cM_k\in\mathsf{PLO}^{\rho}_k$ and set
$\cM_{n+k}=\cM_n\otimes\cM_k\in\mathsf{PLO}^{\rho}_{n+k}$.
For any $\hat\rho^{\,n+k}_{AB}\in\cD_{n+k}(\rho_{AB})$, its first and second block marginals
belong to $\cD_n(\rho_{AB})$ and $\cD_k(\rho_{AB})$, respectively.  If $P_{UV}$ is the
classical distribution induced by $\cM_n\otimes\cM_k$ on $\hat\rho^{\,n+k}_{AB}$, while
$Q_U\otimes Q_V$ is the distribution induced on $\tilde\rho_{AB}^{\otimes(n+k)}$, then
\[
D(P_{UV}\|Q_U\otimes Q_V)
\geq D(P_U\|Q_U)+D(P_V\|Q_V).
\]
Taking the infimum, then the suprema, gives the displayed super-additivity inequality.
\end{remark}
The following theorem obtains a  multi-letter characterization for the Stein's exponent $\theta_{\mathsf{C}}(\epsilon, \rho_{AB}, \tilde \rho_{AB})$ in terms of $\theta^{\star}_{\mathsf{PLO}\text{-}{\rho}}(\rho_{AB},\tilde \rho_{AB})$ and $\theta^{\star}_{\mathsf{BSEP-ZR}}(\rho_{AB},\tilde \rho_{AB})$ defined in \eqref{eq:optimalmultlettexp} and \eqref{eq:limitexp-sepzrexp}, respectively.     
\begin{theorem}[Multi-letter zero-rate Stein's exponent]\label{Thm:Steinsexp-classical-zerorate}
The following hold:
 \begin{enumerate}[label = (\roman*),leftmargin=5.5mm]
\item  Suppose $\rho_{AB}$ and $\tilde \rho_{AB}$ are such that $\rho_A \otimes \rho_B \ll \tilde \rho_{AB}$ and there exist orthonormal eigenbasis $\{\ket{x}\}_{x\in\cX}$ of $\rho_A$ and $\{\ket{\bar x}\}_{\bar x\in\bar\cX}$ of $\rho_B$ for which $[\ket{x\bar x}\!\bra{x\bar x},\tilde\rho_{AB}]=0$ for every $x\in\cX_+$ and $\bar x\in\bar\cX_+$, where $\cX_+$ (resp. $\bar\cX_+$) indexes the eigenvectors of $\rho_A$ (resp. $\rho_B$) with positive eigenvalues. Then
   \begin{align}
\theta_{\mathsf{C}}(\epsilon, \rho_{AB}, \tilde \rho_{AB})=\theta^{\star}_{\mathsf{PLO}\text{-}\rho}(\rho_{AB},\tilde \rho_{AB}), \forall \epsilon \in (0,1). \label{eq:zeroratesteinexp}
   \end{align}
   \item  For arbitrary $\rho_{AB}$ and $\tilde \rho_{AB}$,
    \begin{align}
 \lim_{\epsilon \downarrow 0^+} \theta_{\mathsf{C}}(\epsilon, \rho_{AB}, \tilde \rho_{AB}) =    \theta^{\star}_{\mathsf{BSEP-ZR}}(\rho_{AB},\tilde \rho_{AB}). \label{eq:zerorateweakconv}
\end{align} 
      \end{enumerate} 
\end{theorem} 

Theorem \ref{Thm:Steinsexp-classical-zerorate}$(i)$ provides  a multi-letter characterization of $\theta_{\mathsf{C}}(\epsilon, \rho_{AB}, \tilde \rho_{AB})$ in terms of $\theta^{\star}_{\mathsf{PLO}\text{-}\rho}(\rho_{AB},\tilde \rho_{AB})$,  under the support and product-marginal eigenbasis commutativity conditions stated therein. The achievability of $\theta^{\star}_{\mathsf{PLO}\text{-}\rho}(\rho_{AB},\tilde \rho_{AB})$ follows by considering a scheme that performs a sequence of identical local orthogonal projective measurements in $\mathsf{PLO}^{\rho}_n$  acting on systems $A^n$ and $B^n$ and subsequently applies the optimal  zero-rate classical scheme based on the measurement outcomes. Then, the type II error exponent achieved by this scheme approaches $\theta^{\star}_{\mathsf{PLO}\text{-}\rho}(\rho_{AB},\tilde \rho_{AB})$ in the limit $n \rightarrow \infty$ by optimizing over all such measurements. For the converse direction, we use a bipartite extension of the classical blowing-up lemma (see Lemma \ref{lem:quant-blowup-bipartite} below). 
In the general case, \eqref{eq:zerorateweakconv} shows that  $\theta_{\mathsf{C}}(\epsilon, \rho_{AB}, \tilde \rho_{AB})$   equals  the multi-letter expression $\theta^{\star}_{\mathsf{BSEP-ZR}}(\rho_{AB},\tilde \rho_{AB})$  for vanishing $\epsilon$. The proof of \eqref{eq:zerorateweakconv} relies on extensions of the arguments given in \cite[Theorem 1]{Ahlswede-Csiszar} to the quantum setting. The details are furnished in Section \ref{Sec:Thm:Steinsexp-zerorate-classical-proof}. The second multi-letter expression will be useful for showing later via Example \ref{ex:zeroratecvsq}  that the gap  between the Stein's exponent achievable with zero-rate classical versus quantum communication could be infinite in general. Example \ref{ex:zeroratecvsq} also highlights some implications of Theorem \ref{Thm:Steinsexp-classical-zerorate} to remote state discrimination of so-called data-hiding\cite{Terhal_2001, Eggeling_2002, DiVincenzo_2002, DiVincenzo_2003}, isotropic and Werner states \cite{Werner-1989}. 
  \begin{remark}[Equivalent characterizations under support conditions]\label{Rem:eqprojandbproj}
The supremum and infimum in \eqref{eq:optimalmultlettexp} are achieved (see Appendix \ref{App:supinfach}) when $\rho_A\otimes\rho_B\ll\tilde\rho_{AB}$. Let $\mathsf{BPLO}^{\rho}_n$ denote the binary local PVMs obtained by coarse-graining rank-one PVMs in $\mathsf{PLO}^{\rho}_n$, and define
\begin{align}
\theta^{\star}_{\mathsf{BPLO}\text{-}\rho}(\rho_{AB},\tilde\rho_{AB})
:=\lim_{n\to\infty}\max_{\cM_n\in\mathsf{BPLO}^{\rho}_n}\min_{\hat\rho^n_{AB}\in\cD_n(\rho_{AB})}
\frac{\qrel{\cM_n(\hat\rho^n_{AB})}{\cM_n(\tilde\rho_{AB}^{\otimes n})}}{n}.
\label{eq:limitexp-projlocstein}
\end{align}
Under the assumptions of Theorem~\ref{Thm:Steinsexp-classical-zerorate}(i), Lemma \ref{lem:quant-blowup-bipartite} produces binary projectors in this marginal-compatible class. The converse proof gives
$\theta_{\mathsf C}(\epsilon,\rho_{AB},\tilde\rho_{AB})
=\theta^{\star}_{\mathsf{PLO}\text{-}\rho}(\rho_{AB},\tilde\rho_{AB})
=\theta^{\star}_{\mathsf{BPLO}\text{-}\rho}(\rho_{AB},\tilde\rho_{AB})$ for every $\epsilon\in(0,1)$.
Combining this with the characterization in Theorem~\ref{Thm:Steinsexp-classical-zerorate}(ii), and letting $\epsilon\downarrow0^+$, also gives
$\theta^{\star}_{\mathsf{BSEP-ZR}}(\rho_{AB},\tilde\rho_{AB})=\theta^{\star}_{\mathsf{PLO}\text{-}\rho}(\rho_{AB},\tilde\rho_{AB})$. Hence, under the assumptions of Theorem~\ref{Thm:Steinsexp-classical-zerorate}(i),
\[
\theta_{\mathsf C}(\epsilon, \rho_{AB}, \tilde \rho_{AB})=\theta^{\star}_{\mathsf{BSEP-ZR}}(\rho_{AB}, \tilde \rho_{AB})=\theta^{\star}_{\mathsf{PLO}\text{-}\rho}(\rho_{AB}, \tilde \rho_{AB})=\theta^{\star}_{\mathsf{BPLO}\text{-}\rho}(\rho_{AB}, \tilde \rho_{AB}),~\forall \epsilon \in (0,1). 
\]
\end{remark}
Theorem \ref{Thm:Steinsexp-classical-zerorate} has an interesting corollary when  $\rho_{AB}$ and $\tilde \rho_{AB}$ have the same marginals.
  \begin{cor}[Stein's exponent with same marginals] \label{Cor:zeroSteinexp}
  For any   $\rho_{AB}$ and $\tilde \rho_{AB}$ such that $(i)$ $\tilde \rho_{A}=\rho_{A}$, $\tilde \rho_{B}=\rho_{B}$,  and the conditions in Theorem~\ref{Thm:Steinsexp-classical-zerorate}(i) hold,  or $(ii)$ $\rho_{AB}=\tilde \rho_{AB}$,  we have$\mspace{2 mu}$\footnote{Note that $\rho_{A} \otimes \rho_B$ can have a larger support than $\rho_{AB}$, e.g., when $\rho_{AB}=\Phi$ as given in \eqref{eq:isotropicstate}. Hence, the case $\rho_{AB}=\tilde \rho_{AB}$ has to be considered separately.}  
   \begin{align}
\theta_{\mathsf{C}}(\epsilon, \rho_{AB}, \tilde \rho_{AB}) =0, ~ \forall~\epsilon \in [0,1). \notag 
 \end{align}
\end{cor}
A pertinent question to explore is whether $\theta_{\mathsf{C}}(\epsilon, \rho_{AB}, \tilde \rho_{AB}) $ single-letterizes beyond the instances given in  Corollary \ref{Cor:zeroSteinexp},  given that this is true classically under support conditions. A tempting candidate for the single-letter expression is the quantum analogue of the expression  stated in \eqref{eq:classicaltestscheme}:
\begin{align}
\theta^{\star}_{\mathsf{SL}}(\rho_{AB},\tilde \rho_{AB}):= \min_{\substack{\hat \rho_{AB}:\\ \hat \rho_{A}=\rho_A,~  \hat \rho_{B}=\rho_B} } \qrel{\hat \rho_{AB}}{\tilde \rho_{AB}}.\notag 
\end{align}
  It is easy to see that $\theta^{\star}_{\mathsf{SL}}(\rho_{AB},\tilde \rho_{AB})$ is an upper bound on $\theta^{\star}_{\mathsf{PLO}\text{-}{\rho}}(\rho_{AB},\tilde \rho_{AB})$ from the following:
\begin{align}
\max_{\cM_n \in  \mathsf{PLO}_n\text{-}\rho}  \min_{\hat \rho^n_{AB} \in \cD_n(\rho_{AB})} \frac{\qrel{\cM_n(\hat \rho^n_{AB})}{\cM_n(\tilde \rho_{AB}^{\otimes n})}}{n} 
& \leq \max_{\cM_n \in  \mathsf{PLO}_n} \min_{\substack{\hat \rho^{\otimes n}_{AB}:\\ \hat \rho_A=\rho_A,\hat \rho_B=\rho_B}} \frac{\qrel{\cM_n(\hat \rho_{AB}^{\otimes n})}{\cM_n(\tilde \rho_{AB}^{\otimes n})}}{n} \notag \\
& \leq   \min_{\substack{\hat \rho_{AB}:\\ \hat \rho_A=\rho_A,\hat \rho_B=\rho_B}} \max_{\cM_n \in  \mathsf{ALL_n}}\frac{\qrel{\cM_n(\hat \rho_{AB}^{\otimes n})}{\cM_n(\tilde \rho_{AB}^{\otimes n})}}{n}, \notag 
\end{align}
where  the penultimate inequality is due to $\min_{a \in \cA} \max_{b \in \cB} f(a,b) \geq  \max_{b \in \cB} \min_{a \in \cA} f(a,b)$ for any function $f:\cA \times \cB \rightarrow \RR$, and the fact that $\mathsf{PLO}_n$ is a subset of the set of all measurements $\mathsf{ALL_n}$. The claim $\theta^{\star}_{\mathsf{PLO}\text{-}{\rho}}(\rho_{AB},\tilde \rho_{AB}) \leq \theta^{\star}_{\mathsf{SL}}(\rho_{AB},\tilde \rho_{AB})$ then follows by taking limits $n\rightarrow \infty$ and noting that regularized measured relative entropy optimized over all measurements asymptotically achieves quantum relative entropy (see, e.g., \cite{tomamichel2015quantum}). Consequently, Theorem \ref{Thm:Steinsexp-classical-zerorate}$(i)$ yields that $\theta^{\star}_{\mathsf{SL}}(\rho_{AB},\tilde \rho_{AB})$ is also an upper bound on $\theta_{\mathsf{C}}(\epsilon, \rho_{AB}, \tilde \rho_{AB})$ when the support and commutativity conditions in Theorem~\ref{Thm:Steinsexp-classical-zerorate}(i) holds. In addition, since $\theta^{\star}_{\mathsf{SL}}(\rho_{AB},\tilde \rho_{A} \otimes \tilde \rho_{B})=\qrel{\rho_{A}}{\tilde \rho_A}+\qrel{\rho_{B}}{\tilde \rho_B}$, Theorem~\ref{Thm:Steinsexp-zerorate} shows that $ \theta(\epsilon, \rho_{AB}, \tilde \rho_{A} \otimes \tilde \rho_{B}) =\theta^{\star}_{\mathsf{SL}}(\rho_{AB},\tilde \rho_{A} \otimes \tilde \rho_{B})$ for $\epsilon \in (0,1)$.

To shed more light into the aspect of single-letterization of the Stein's exponent, we next investigate the case when the underlying states for hypothesis testing are CQ states $\rho_{XB}$ and $\tilde\rho_{XB}$. 
\begin{cor}[Stein's exponent for CQ states]\label{Cor:Steinexpcqstates}
 Let $\rho_{XB}$ and $\tilde \rho_{XB}$ be CQ states of the form
 \begin{equation}
       \label{eq:CQstatesex}\begin{aligned}  \rho_{XB}&=\sum_{x \in \cX} p_{X}(x) \ketbra{x}{x} \otimes \rho_x^B,  \\
        \tilde \rho_{XB}&=\sum_{x \in \cX} \tilde p_{X}(x) \ketbra{x}{x} \otimes \tilde \rho_x^B, 
    \end{aligned}
    \end{equation}
   where $p_X, \tilde p_X$ are probability mass functions and $\rho_x^B$, $\tilde \rho_x^B$ are density operators for all $x \in \cX$. 
    Then, we have 
    \begin{subequations}
    \begin{align} 
        \theta_{\mathsf{C}}(\epsilon, \rho_{XB}, \tilde \rho_{XB})\geq\theta^{\star}_{\mathsf{CQ}}(\rho_{XB},\tilde \rho_{XB})\geq\theta^{\star}_{\mathsf{CQ-BP}}(\rho_{XB},\tilde \rho_{XB}),~ \forall~ \epsilon \in (0,1],  \label{eq:SteinexpCQchar}
    \end{align}
 \text{where }
    \begin{align}
\theta^{\star}_{\mathsf{CQ}}(\rho_{XB},\tilde \rho_{XB})  &:=  \lim_{n \rightarrow \infty}\max_{ \cP_n \in \mathsf{P}_n^{\rho_B}(B^n)}   \min_{\hat \rho^n_{XB} \in \tilde{\cD}_n(\rho_{XB}) }\frac{\qrel{(I_{X^n} \otimes \cP_n)(\hat \rho^n_{XB})}{(I_{X^n} \otimes \cP_n)(\tilde \rho^{\otimes n}_{XB})}}{n}, \label{eq:steinexpcq}\\
\theta^{\star}_{\mathsf{CQ-BP}}(\rho_{XB},\tilde \rho_{XB})&:=\lim_{n \rightarrow \infty}\max_{\cP_n \in \mathsf{BP}_n^{\rho_B}(B^n)}   \min_{\hat \rho^n_{XB} \in \tilde{\cD}_n(\rho_{XB}) }\frac{\qrel{(I_{X^n} \otimes \cP_n)(\hat \rho^n_{XB})}{(I_{X^n} \otimes \cP_n)(\tilde \rho^{\otimes n}_{XB})}}{n}, \label{eq:steinexpcqbo}\\
\tilde{\cD}_n(\rho_{XB}) &:=\left\{\hat \rho^n_{XB}=\sum_{x^n \in \cX^n} p_{X^n}(x^n) \ket{x^n}\bra{x^n} \otimes \hat \rho_{x^n}^{B^n} :\hat \rho^n_{B}= \sum_{x^n \in \cX^n} p_{X^n}(x^n) \hat \rho_{x^n}^{B^n} = \rho_{B}^{\otimes n} \right\}, \notag
    \end{align}     
    \end{subequations}
$p_{X^n}(x^n):=\prod_{i=1}^n p_{X_i}(x_i)$. Here $\mathsf{P}_n^{\rho_B}(B^n)$ denotes the set of orthogonal rank-one PVMs $\cP_n=\{P_{y_n}\}_{y_n\in[1:d_B^n]}$ satisfying $[P_{y_n},\rho_B^{\otimes n}]=0$ for every $y_n$, and $\mathsf{BP}_n^{\rho_B}(B^n)$ denotes their binary coarse-grainings
\[
\left\{\sum_{y_n\in\hat\cY_n}P_{y_n},\ I_{B^n}-\sum_{y_n\in\hat\cY_n}P_{y_n}\right\}.
\]
Additionally, if  $\rho_{XB}$ and $\tilde \rho_{XB}$ satisfy the support and product-marginal eigenbasis commutativity conditions in Theorem \ref{Thm:Steinsexp-classical-zerorate}$(i)$, then the inequalities in \eqref{eq:SteinexpCQchar} hold with equality. 
\end{cor}
The proof of Corollary \ref{Cor:Steinexpcqstates} is given in Section \ref{Sec:Cor:Steinexpcqstates-proof} and   essentially relies on the structure of CQ states and Lemma~\ref{Lem:qustwithclmarg}(ii) to show \eqref{eq:SteinexpCQchar}. The matching upper bound on $\theta_{\mathsf{C}}(\epsilon, \rho_{XB}, \tilde \rho_{XB})$ required for equality in \eqref{eq:SteinexpCQchar} is then obtained by invoking the bipartite generalization of the blowing-up lemma.

The next proposition provides an upper bound on $\theta_{\mathsf{C}}(\epsilon, \rho_{XB}, \tilde \rho_{XB})$ for specific instances of CQ states, which will be useful to highlight the gap between    $\theta_{\mathsf{C}}(\epsilon, \rho_{XB}, \tilde \rho_{XB})$ and $\theta^{\star}_{\mathsf{SL}}(\rho_{XB},\tilde \rho_{XB})$.
\begin{prop}[Stein's exponent and single-letter formula]\label{Prop:counterexcqstate}
Consider  CQ states of the form given in \eqref{eq:CQstatesex}  with $\cX=\{0,1\}$,  $p_{X}(0)=p_{X}(1)=0.5$,   $0<\tilde p_{X}(0) <1$,
$\rho_0=\rho_1=\tau$ being an arbitrary qubit density operator, and  $\tilde \rho_0, \tilde \rho_1$ being qubit density operators of full rank. Then
    \begin{align}
\theta_{\mathsf{C}}(\epsilon, \rho_{XB}, \tilde \rho_{XB}) \leq \qrel{p_X}{\tilde p_X}+\qrel{\tau}{\omega(\tilde\rho_0,\tilde\rho_1)}, \label{eq:steinexpgap}
    \end{align}
    where $\omega(\sigma_0,\sigma_1):= \sigma_0^{1/2}\big(\sigma_0^{-1/2} \sigma_1 \sigma_0^{-1/2}\big)^{1/2}\sigma_0^{1/2}$ is the  Kubo-Ando geometric mean \cite{Kubo1979} between $\sigma_0$ and $\sigma_1$.  
   In particular, when $\tau=\psi$ with $\psi$ denoting a  pure state,
    \begin{align}
\theta_{\mathsf{C}}(\epsilon, \rho_{XB}, \tilde \rho_{XB}) \leq \qrel{p_X}{\tilde p_X}+\qrel{\psi}{\omega(\tilde\rho_0,\tilde\rho_1)} = \theta^{\star}_{\mathsf{SL}}(\rho_{XB},\tilde \rho_{XB})-\kappa(\psi,\tilde \rho_0,\tilde \rho_1), \label{eq:steinexpgapspec}
    \end{align}   
    where 
\begin{subequations}\label{eq:gapcqstatesinglet}
   \begin{align}   \theta^{\star}_{\mathsf{SL}}(\rho_{XB},\tilde \rho_{XB})&:= \min_{\substack{\hat \rho_{XB}=\sum_{x \in \cX} \hat p_{X}(x) \ket{x}\bra{x} \otimes \hat \rho_x^B:\\\hat \rho_{X}=\rho_X,~  \hat \rho_{B}=\rho_B} } \qrel{\hat \rho_{XB}}{\tilde \rho_{XB}}=\qrel{p_X}{\tilde p_X}+ \frac{1}{2} \big(\qrel{\psi}{\tilde \rho_0}+ \qrel{\psi}{\tilde \rho_1}\mspace{-2 mu}\big), \label{eq:cqstsinglet} \\
\kappa(\psi,\tilde \rho_0,\tilde \rho_1)&:= \frac 12 \mspace{2 mu} \Big(\qrel{\psi}{\tilde \rho_0}+ \qrel{\psi}{\tilde \rho_1}-2\, \qrel{\psi}{\omega(\tilde \rho_0 ,\tilde \rho_1 )}\mspace{-4 mu}\Big).\label{eq:gapcqstatesinglet1}
    \end{align}  
    \end{subequations} 
\end{prop}
 The proof of Proposition \ref{Prop:counterexcqstate} is given in Section \ref{Sec:Prop:counterexcqstate-proof}. The upper bound in \eqref{eq:steinexpgap} is established by a direct converse that combines a classical blowing-up argument on the binary $X^n$ system, the Kubo-Ando geometric-mean inequality \cite[Theorem 3]{Ando-concavity} , and a binary-test bound based on the sandwiched R\'enyi divergence on the $B^n$ system. 
 
 Equation \eqref{eq:steinexpgapspec} shows  that there is a strict gap between  $\theta^{\star}_{\mathsf{SL}}(\rho_{XB},\tilde \rho_{XB})$ and $\theta_{\mathsf{C}}(\epsilon, \rho_{XB}, \tilde \rho_{XB})$ for the CQ states in Proposition \ref{Prop:counterexcqstate} with $\tau=\psi$ when $\kappa(\psi,\tilde \rho_0,\tilde \rho_1)>0$. Indeed, with $\ket{+}=(\ket{0}+\ket{1})/\sqrt{2}$,  $\ket{-}=(\ket{0}-\ket{1})/\sqrt{2}$  and computing $\kappa(\psi,\tilde \rho_0,\tilde \rho_1)$ numerically for 
 \begin{align}
 \psi=\ket{0}\bra{0}, \quad \tilde \rho_0=(1-p) \ket{0}\bra{0}+p\ket{1}\bra{1}, \quad \tilde \rho_1=(1-q) \ket{+}\bra{+}+q\ket{-}\bra{-}, \quad \mbox{with } p=0.6 \mbox{ and } q=0.9, \notag   
 \end{align}
 shows that  $\kappa(\psi,\tilde \rho_0,\tilde \rho_1)\approx 0.0178$.     
 Hence, the single-letter expression given in \eqref{eq:cqstsinglet} does not  characterize the  Stein's exponent for this case.

\subsection{Stein's exponent with classical versus quantum communication}
It is known classically  that $\theta_{\mathsf{ZRC}}^{\star}(p_{XY}, \tilde p_{XY})$ given in \eqref{eq:classicaltestscheme} does not characterize the  Stein's exponent in general, when the support condition $\tilde p_{XY}>0$ is violated.  In fact, the gap can be infinite  when $p_{XY}$ and $\tilde p_{XY}$ have orthogonal support (see \cite[Remark 2]{Han-1987}).  Similar phenomenon also extends to the quantum setting. In fact,  $\theta(\epsilon, \rho_{AB}, \tilde \rho_{AB})$ and $\theta_{\mathsf{C}}(\epsilon, \rho_{AB}, \tilde \rho_{AB})$ can exceed $\theta^{\star}_{\mathsf{SL}}(\rho_{AB},\tilde \rho_{AB})$$\big(\mspace{-3 mu}\geq \theta^{\star}_{\mathsf{PLO}\text{-}{\rho}}(\rho_{AB},\tilde \rho_{AB})\big)$ by an infinite amount,  as the next example highlights.
\begin{example}[Violation of single-letter upper bound]\label{ex:steinexpinfin}
Let 
\begin{align}
   \rho_{AB}&= \frac{(\ket{00}+\ket{11})(\bra{00}+\bra{11})}{2} \quad 
 \mbox{ and } \quad 
   \tilde \rho_{AB}= \frac{(\ket{01}+\ket{10})(\bra{01}+\bra{10})}{2}. \notag
\end{align}
Consider the scheme where Alice and Bob performs the local measurement $ \big\{P_x^A\big\}_{x \in \{0,1\}} \otimes \big\{P_y^B\big\}_{y \in \{0,1\}} \in \mathsf{PLO} $, where $P_0=\ket{0}\bra{0}$ and $ P_1=\ket{1}\bra{1}$, and communicates the one-bit outcome $x$ and $y$  to Charlie, respectively. Charlie decides in favour of $H_0$ if $x=y$, and $H_1$ otherwise. Then, since $x=y$ holds under the null and $x\neq y$ holds under the alternative, $H_0$ and $H_1$ can be perfectly discriminated. Hence,   $\theta(\epsilon, \rho_{AB}, \tilde \rho_{AB}) =\theta_{\mathsf{C}}(\epsilon, \rho_{AB}, \tilde \rho_{AB})=\infty$ for all $\epsilon \in [0,1]$. On the other hand, $\theta^{\star}_{\mathsf{SL}}(\rho_{AB},\tilde \rho_{AB})=0$ since $\rho_A=\tilde \rho_A$ and $\rho_B=\tilde \rho_B$. Yet another example exhibiting this phenomenon is $\rho_{AB}= (\ket{++}+\ket{--})(\bra{++}+\bra{--})/2$ and  $\tilde \rho_{AB}= (\ket{+-}+\ket{-+})(\bra{+-}+\bra{-+})/2$.    In other words, when $\rho_{AB}$ and $\tilde \rho_{AB}$ have orthogonal support, $\theta(\epsilon, \rho_{AB}, \tilde \rho_{AB})$ and $\theta_{\mathsf{C}}(\epsilon, \rho_{AB}, \tilde \rho_{AB})$ can be strictly greater than  $\theta^{\star}_{\mathsf{SL}}(\rho_{AB},\tilde \rho_{AB})$.
\end{example}
Next, we show that the Stein's exponents $\theta(\epsilon, \rho_{AB}, \tilde \rho_{AB})$ and $\theta_{\mathsf{C}}(\epsilon, \rho_{AB}, \tilde \rho_{AB})$ does not coincide in general. To this end, it will be useful to obtain an upper bound on $\theta_{\mathsf{C}}(\epsilon, \rho_{AB}, \tilde \rho_{AB})$ (for  the case of vanishing $\epsilon$), when $\rho_{AB}$ and $\tilde \rho_{AB}$ are isotropic or  Werner states violating the support condition.   A general isotropic state can be characterized completely by a single parameter \(p\in[0,1]\) as 
\begin{equation}
    \dutchcal{i}(p) := p \Phi + (1-p) \Phi^\perp, \label{eq:isotropic}
\end{equation}
where \(\Phi\) denotes the maximally entangled state and \(\Phi^\perp\) is its orthogonal complement. Using the canonical basis \(\{\ket{i}\}_{i=1}^d\) on \(A\) and \(B\), respectively, the extremal isotropic states are given by
\begin{align}
    \Phi &= \frac{1}{d} \sum_{i,j =1}^d \ketbra{i}{j}_A \otimes \ketbra{i}{j}_B \qquad  \text{and} \qquad  \Phi^\perp = \frac{I_{AB} - \Phi}{d^2 -1}. \label{eq:isotropicstate}
\end{align}  
A general Werner state \cite{Werner-1989} is given by
\begin{equation}
    \dutchcal{w}(p) := p \Theta + (1-p) \Theta^\perp \, , \label{eq:werner}
\end{equation}
where \(\Theta\) is the completely symmetric and \(\Theta^\perp\) the completely anti-symmetric state. These states are defined as
\begin{align}
    \Theta &:= \frac{I_{AB} + F}{d(d+1)} \qquad  \text{and} \qquad  \Theta^\perp := \frac{I_{AB} - F}{d(d-1)} \,, \notag
\end{align}
with the swap operator \(F := \sum_{i,j} \ketbra{i}{j}_A \otimes \ketbra{j}{i}_B \). 
\begin{cor}[Stein's exponent without support condition]\label{Cor:upperbndsteinexp}
For any $p \in [0,1]$, 
 \begin{align}
& 0 \leq \lim_{\epsilon \downarrow 0^+} \theta_{\mathsf{C}}(\epsilon,\dutchcal{i}(p), \Phi^\perp) \leq  \log (pd+1), \notag \\
& 0 \leq \lim_{\epsilon \downarrow 0^+} \theta_{\mathsf{C}}(\epsilon,\dutchcal{w}(p), \Theta) \leq  \log \left(\frac{d+1-2p}{d-1}\right). \notag
\end{align} 
\end{cor}
The proof of the above corollary follows from \eqref{eq:zerorateweakconv}  by using the results in \cite{RSB-IT-2025} to upper bound $\theta^{\star}_{\mathsf{BSEP-ZR}}(\cdot,\cdot)$.

The next example shows that $\theta(\epsilon, \rho_{AB}, \tilde \rho_{AB})$  can be strictly larger than $\theta_{\mathsf{C}}(\epsilon, \rho_{AB}, \tilde \rho_{AB})$.   More radically,  even if both encoders at Alice and Bob are allowed to communicate just one qudit to Charlie, the gap in performance between the two scenarios can be infinite.
\begin{example}[Zero-rate classical versus quantum communication] \label{ex:zeroratecvsq}
Consider the case of testing between two orthogonal isotropic states $\rho_{AB}=\Phi$ and $\tilde \rho_{AB}=\Phi^{\perp}$ shared between Alice and Bob under the two hypotheses. Then, Alice and Bob can transmit their entire $d$-dimensional local systems (one $d$-level qudit each) to Charlie and achieve a Stein's exponent\,\footnote{More generally, if Alice and Bob transmit their entire local quantum systems to Charlie, any two orthogonal bipartite states can be perfectly distinguished, and hence the optimal error-exponent is infinite.} $\theta(\epsilon,\Phi, \Phi^{\perp})=\infty$  for all $\epsilon \in [0,1]$. This is because Charlie having access to $A$ and $B$ can perfectly distinguish  these two orthogonal states on them via the measurement $\cM=\{\Phi, \Phi^{\perp}\}$.  On the other hand, by Corollary \ref{Cor:upperbndsteinexp}, $\lim_{\epsilon \downarrow 0^+}\theta_{\mathsf{C}}(\epsilon,\Phi, \Phi^{\perp}) \leq \log (d+1)$. Hence, zero-rate quantum communication can strictly  improve the   error-exponent achievable by zero-rate classical communication in this case. Similar remark also applies for $\rho_{AB}=\Theta^{\perp}$ and $\tilde \rho_{AB}=\Theta$, where $\lim_{\epsilon \downarrow 0^+}\theta_{\mathsf{C}}(\epsilon,\Theta^{\perp}, \Theta) \leq  \log \big(d+1/(d-1)\big)$ while $\theta(\epsilon,\Theta^{\perp}, \Theta) =\infty$.   
\end{example}
\section{Proofs}\label{Sec:Proofs}
This section contains proofs of the results stated in Section \ref{Sec:mainres}. 
For ease of readability, longer proofs  are split into lemmas, whose  proofs are presented later in the order of their numbering. 
\subsection{Proof of Theorem \ref{Thm:Steinsexp-zerorate}}\label{Sec:Thm:Steinsexp-zerorate-proof}
\noindent \textbf{Achievability:} Here, we show achievability of  Stein's exponent given in \eqref{eq: steinexp-productalt}  via a direct argument using  relatively typical projections, i.e., 
\begin{align}
   \theta(\epsilon, \rho_{AB}, \tilde \rho_{A} \otimes \tilde \rho_{B})  \geq \qrel{\rho_{A}}{\tilde \rho_A}+\qrel{\rho_{B}}{\tilde \rho_B}, ~\forall \epsilon \in (0,1]. \notag
\end{align}
First, consider the case when $\rho_A\not\ll\tilde\rho_A$ or $\rho_B\not\ll\tilde\rho_B$. Assume that the former holds, and  let $\Pi_{\tilde\rho_A}$ denote the support projection of $\tilde\rho_A$.  Set
\[
Q_A:=I_A-\Pi_{\tilde\rho_A},\qquad
p_A:=\tr{\rho_AQ_A}>0.
\]
On $n$ copies, Alice performs the binary measurement
\[
\left\{T_{A,n},\,I_{A^n}-T_{A,n}\right\},\qquad
T_{A,n}:=I_{A^n}-(I_A-Q_A)^{\otimes n},
\]
and communicates its one-bit outcome to Charlie, who decides in favour of the null hypothesis whenever the
$T_{A,n}$ outcome occurs.  Under the null hypothesis,
\[
\tr{\rho_A^{\otimes n}T_{A,n}}
=1-(1-p_A)^n\longrightarrow1,
\]
whereas, since $Q_A\tilde\rho_A=0$,
\[
\tr{\tilde\rho_A^{\otimes n}T_{A,n}}=0.
\]
Hence, the type-I error probability tends to zero while the type-II error probability is identically zero. Therefore,  the achievable
Stein's exponent is infinite for every fixed $\epsilon \in (0,1]$, and therefore, in agreement with
$D(\rho_A\|\tilde\rho_A)=\infty$.  The same argument applies if
$\rho_B\not\ll\tilde\rho_B$.  

Next, consider that $\rho_A\ll\tilde\rho_A$ and $ \rho_B\ll\tilde\rho_B$ hold. For $\delta >0$, let $ P_{\delta,n}(\rho,\sigma):=\sum_{x^n \in \cA_n(\delta,\rho,\sigma)} P_{x^n}$ denote the orthogonal projection considered in \cite[Equation 10]{Bjela-Schultze-2012}, where $P_{x^n}=\ket{x^n}\bra{x^n}$ for an orthonormal eigenvector $\ket{x^n}$ of $\sigma^{\otimes n}$ and 
\begin{align}
   \cA_n(\delta,\rho,\sigma)&:=\{x^n: e^{n(\tr{\rho \log \sigma}-\delta)} \leq \tr{\sigma^{\otimes n} P_{x^n}}    \leq e^{n(\tr{\rho \log \sigma}+\delta)}\}. \notag
\end{align}
Consider
\begin{align}
    \bar{M}_{\delta,n}^{A^nB^n}(\rho_{AB},\tilde \rho_{AB})=  \hat{M}_{\delta,n}^{A^n}(\rho_A,\tilde \rho_A) \otimes \hat{M}_{\delta,n}^{B^n}(\rho_B,\tilde \rho_B), \notag
\end{align}
where for density operators $\rho$ and $\sigma$,
\begin{align}
    \hat M_{\delta,n}(\rho,\sigma)=P_{\delta,n}(\rho,\sigma) P_{\delta,n}(\rho,\rho) P_{\delta,n}(\rho,\sigma). \notag
\end{align}
 Alice and Bob first performs local measurements using  binary outcome  measurements 
 \begin{align}
    & \cM_n^{A^n}=\{M_0^{A^n},M_1^{A^n}\}:=\big\{\hat M_{\delta,n}^{A^n}(\rho_A,\tilde \rho_A),I_{A^n}-\hat M_{\delta,n}^{A^n}(\rho_A,\tilde \rho_A)\big\}, \notag \\
 \mbox{and} \quad  & \cM_n^{B^n}=\{M_0^{B^n},M_1^{B^n}\}:=\big\{\hat M_{\delta,n}^{B^n}(\rho_B,\tilde \rho_B),I_{B^n}-\hat M_{\delta,n}^{B^n}(\rho_B,\tilde \rho_B)\big\}, \notag
 \end{align} respectively. The  outcomes are sent to Charlie who decides in favour of the null hypothesis if and only if both outcomes are zero. We next evaluate the type I and type II error probability achieved by this one-bit communication scheme. 
We have using \eqref{eq:probintersecevent} that
\begin{align}
    \tr{\bar{M}_{\delta,n}^{A^nB^n}(\rho_{AB},\tilde \rho_{AB}) \rho^{\otimes n}_{AB}} &\geq  \tr{\big(\hat M_{\delta,n}^{A^n}(\rho_A,\tilde \rho_A) \otimes I^{B^n}\big) \rho^{\otimes n}_{AB}}+\tr{\big(I^{A^n}\otimes \hat M_{\delta,n}^{B^n}(\rho_B,\tilde \rho_B)  \big)\rho^{\otimes n}_{AB}}-1 \notag \\
    &= \tr{\hat M_{\delta,n}^{A^n}(\rho_A,\tilde \rho_A) \rho_{A}^{\otimes n}}+ \tr{ \hat M_{\delta,n}^{B^n}(\rho_B,\tilde \rho_B) \rho_{B}^{\otimes n}}-1 \notag \\
    & \rightarrow 1, \notag
\end{align}
since $\tr{\hat M_{\delta,n}^{A^n}(\rho_A,\tilde \rho_A) \rho_{A}^{\otimes n}} \rightarrow 1$ and $\tr{\hat M_{\delta,n}^{B^n}(\rho_B,\tilde \rho_B) \rho_{B}^{\otimes n}} \rightarrow 1$ based on  \cite{Bjela-Schultze-2012}[Lemma 4 (4)]. Hence, the type I error probability vanishes asymptotically. Moreover, the type II error probability can be upper bounded as 
\begin{align}
      \tr{\bar{M}_{\delta,n}^{A^nB^n}(\rho_{AB},\tilde \rho_{AB})(\tilde  \rho_{A}^{\otimes n} \otimes \tilde  \rho_{B}^{\otimes n})}&=\tr{\hat{M}_{\delta,n}^{A^n}(\rho_{A},\tilde \rho_{A})\tilde  \rho_{A}^{\otimes n}} \tr{\hat{M}_{\delta,n}^{B^n}(\rho_{B},\tilde \rho_{B}) \tilde  \rho_{B}^{\otimes n}}  \notag \\
      & \leq  e^{-n\big(\qrel{\rho_A}{\tilde \rho_A}+\qrel{\rho_B}{\tilde \rho_B}-4\delta\big)}, \notag
\end{align}
where the last inequality follows due to \cite{Bjela-Schultze-2012}[Equation 20].
This implies that $\theta(\epsilon, \rho_{AB}, \tilde \rho_{AB}) \geq \qrel{\rho_A}{\tilde \rho_A}+\qrel{\rho_B}{\tilde \rho_B}-4\delta$ for all $\epsilon \in (0,1]$. Since 
$\delta>0$ is arbitrary, the desired  claim follows.

\noindent \textbf{Converse:}
To complete the proof of \eqref{eq: steinexp-productalt}, we need to establish the converse, i.e., 
    \begin{align}
  \theta(\epsilon, \rho_{AB}, \tilde \rho_{A} \otimes \tilde \rho_{B})   & \leq \qrel{\rho_{A}}{\tilde \rho_A}+\qrel{\rho_{B}}{\tilde \rho_B},\quad \forall~\epsilon \in [0,1).\label{eq:quantconv}
   \end{align}
The above inequality will be a straightforward consequence of  the following non-asymptotic second order converse. 
\begin{lemma}[Non-asymptotic strong converse via reverse hypercontractivity] \label{Lem:strong-converse-revhyp}
Consider any $\rho_{AB}$ and $\tilde \rho_{AB}=\tilde \rho_A \otimes \tilde \rho_B$. Then, 
for any CPTP maps $\cF_n$, $\cG_n,\cT_n$ such that $ \alpha_n(\cF_n,\cG_n,\cT_n) \leq \epsilon < 1$, we have
   \begin{align}
   -\frac{1}{n}\log \beta_n(\cF_n,\cG_n,\cT_n) &\leq  \qrel{\rho_A}{\tilde \rho_A}+\qrel{\rho_B}{\tilde \rho_B}+2 \sqrt{  \frac{\big(\zeta(\rho_A,\tilde \rho_A)+\zeta(\rho_B,\tilde \rho_B)\big)\xi(|W_n|,\epsilon)}{n}}+\frac{\xi(|W_n|,\epsilon)}{n} , \label{eq:type2err-strongconv-hypcont}
\end{align}
where $\xi(|W_n|,\epsilon) \coloneqq 2\log \left(\frac{|W_n|^2}{1-\epsilon}\right)$, and $\zeta(\rho,\sigma)\coloneqq \big\|\sigma^{-\frac 12}\rho \sigma^{-\frac 12}\big\|_{\infty}$ when $\rho \ll \sigma$ (with $\sigma^{-1/2}$ denoting the generalized inverse on $\operatorname{supp}(\sigma)$) and $\zeta(\rho,\sigma)=\infty$ otherwise.
\end{lemma} 
The proof of Lemma \ref{Lem:strong-converse-revhyp} is given in Section \ref{Sec:Lem:strong-converse-revhyp-proof} and  relies on a reverse hypercontractivity result for the generalized depolarizing semigroup proved in \cite{Beigi-2020}.

\medskip

Taking supremum over all $\cF_n$, $\cG_n,\cT_n$ such that $ \alpha_n(\cF_n,\cG_n,\cT_n) \leq \epsilon $ in \eqref{eq:type2err-strongconv-hypcont} followed by limit $n \rightarrow \infty$, we obtain \eqref{eq:quantconv} since $\xi(|W_n|,\epsilon)=o(n)$ (due to $\log |W_n|=o(n)$) in the zero-rate setting. This concludes the proof of Theorem \ref{Thm:Steinsexp-zerorate}.  
\subsection{Proof of Theorem \ref{Thm:Steinsexp-classical-zerorate}}\label{Sec:Thm:Steinsexp-zerorate-classical-proof}
\noindent \textbf{Part $(i)-$Achievability:} To prove the achievability part of \eqref{eq: steinexp-productalt}, we will show a more general result, namely that for any $\rho_{AB}$ and $\tilde \rho_{AB}$,
\begin{align}
    \theta_{\mathsf{C}}(\epsilon, \rho_{AB}, \tilde \rho_{AB})  
   &\geq \theta^{\star}_{\mathsf{PLO}\text{-}\rho}(\rho_{AB},\tilde \rho_{AB}), ~ \forall \epsilon \in (0,1]. \label{eq:achievabilityzerorate}
\end{align}
We will derive \eqref{eq:achievabilityzerorate} by considering a scheme that communicates just one bit of information, each from Alice and Bob to Charlie. Note that this is an achievable scheme for zero-rate classical or quantum communication from Alice and Bob to Charlie  as one bit can be encoded into a qubit.   
    The scheme consists of applying a sequence of identical composite measurements, each acting locally on $A^n$ and $B^n$, and applying the optimal classical  zero-rate scheme on the i.i.d. outcomes generated by the measurements.   
Fix a local measurement $\cM_n=\cP_n^{A^n} \otimes \cP_n^{B^n} \in  \mathsf{PLO}^{\rho}_n $, where we recall that $\cP_n^{A^n}=\{P_{x}=\ket{x}\bra{x}\}_{x \in [1:d_A^n]}$ and $\cP_n^{B^n}=\{P_{y}=\ket{y}\bra{y}\}_{y \in [1:d_B^n]}$ are orthogonal rank-one projections on $\HH_{d_A^n}$ and $\HH_{d_B^n}$. Let  
\begin{align}
& p_{XY}^{(n)}(x,y)=\tr{\big(P_{x}^{A^n} \otimes P_{y}^{B^n}\big) \rho_{AB}^{\otimes n}} , \notag \\
& \tilde p_{XY}^{(n)}(x,y)=\tr{\big(P_{x}^{A^n} \otimes P_{y}^{B^n}\big) \tilde \rho_{AB}^{\otimes n}}, \notag  
\end{align}
  be the p.m.f.'s of the measurement outcomes under the null and alternative hypothesis, respectively. Let $(\hat x_1,\ldots,\hat x_K) $ be the i.i.d. measurement outcomes obtained by applying  the measurement $\bigotimes_{k=1}^K \{P_{x_k}^{A^n}\}_{x_k \in [1:d_A^n]}$ on the system $(A^{n})^{K}$, and similarly $(\hat y_1,\ldots,\hat y_K) $ be the i.i.d. measurement outcomes obtained by applying  the measurement $\bigotimes_{k=1}^K \{P_{y_k}^{B^n}\}_{y_k \in [1:d_B^n]}$ on the system $(B^{n})^{K}$. We next apply the optimal classical testing scheme leading to \eqref{eq:classicaltestscheme},  with $(p_{XY},\tilde p_{XY})$ replaced by $(p_{XY}^{(n)},\tilde p_{XY}^{(n)})$. Denoting by  $(\cF_{nk},\cG_{nk}, \cT_{nk})$ the resulting  one-bit (zero-rate) communication scheme, we obtain that
\begin{align}
\liminf_{k \rightarrow \infty}- \frac{1}{nk}\log \beta_{nk}(\cF_{nk},\cG_{nk}, \cT_{nk}) & \geq \min_{\substack{\hat p_{XY}^{(n)}: \\ \hat p_{X}^{(n)}=p_{X}^{(n)},~ \hat p_{Y}^{(n)}=p_{Y}^{(n)}} } \frac{\mathsf{D}\big(\hat p_{XY}^{(n)}\|\tilde p_{XY}^{(n)}\big)}{n} \notag \\
& \geq \min_{\hat \rho^n_{AB} \in \cD_n(\rho_{AB})} \frac{\qrel{\cM_n(\hat \rho^n_{AB})}{\cM_n(\tilde \rho_{AB}^{\otimes n})}}{n}. \notag
\end{align}
In the above, the final inequality follows since $\cM_n(\tilde \rho_{AB}^{\otimes n})=\sum_{(x,y) \in \cX_n \times \cY_n} \tilde  p_{XY}^{(n)}(x,y) \ket{xy}\bra{xy}$, and  by Lemma \ref{Lem:qustwithclmarg} $(i)$ given below, for every $\cM_n=\cP_n^{A^n} \otimes \cP_n^{B^n} \in  \mathsf{PLO}^{\rho}_n$ and $\hat p_{XY}^{(n)}$ such that $ \hat p_{X}^{(n)}=p_{X}^{(n)}$, $\hat p_{Y}^{(n)}=p_{Y}^{(n)}$, there exists $\hat \rho^n_{AB} \in \cD_n(\rho_{AB})$ (see \eqref{eq:minimizset}) such that 
\begin{align}
\cM_n(\hat \rho^n_{AB})=    \sum_{(x,y) \in \cX_n \times \cY_n} \hat p_{XY}^{(n)}(x,y) \ket{xy}\bra{xy}. \notag 
\end{align}
Taking the supremum over all $\cM_n \in \mathsf{PLO}^{\rho}_n$ and $n \in \NN$  yields \eqref{eq:achievabilityzerorate} for every $\epsilon \in (0,1]$. 
\begin{lemma}[Quantum lifting for marginal-compatible measurements]\label{Lem:qustwithclmarg}
Let $\rho_{AB}$ be a bipartite state and let $\cM=\cP^A\otimes\cP^B$ be a local rank-one PVM with $\cP^A=\{P_x^A\}_{x\in\cX}$ and $\cP^B=\{P_y^B\}_{y\in\cY}$ satisfying
\[
[P_x^A,\rho_A]=0,\qquad [P_y^B,\rho_B]=0\qquad\text{for all }(x,y) \in \cX \times \cY.
\]
Set $p_X(x)=\tr{P_x^A\rho_A}$ and $p_Y(y)=\tr{P_y^B\rho_B}$. For every distribution $\hat p_{XY}$ with $\hat p_X=p_X$ and $\hat p_Y=p_Y$, there exists a state $\hat\rho_{AB}$ such that
\[
\hat\rho_A=\rho_A,\quad \hat\rho_B=\rho_B,\quad \mbox{and} \quad 
\tr{(P_x^A\otimes P_y^B)\hat\rho_{AB}}=\hat p_{XY}(x,y).
\]
In particular, for rank-one marginal eigenbasis PVMs one may take
\[
\hat\rho_{AB}=\sum_{x,y}\hat p_{XY}(x,y)P_x^A\otimes P_y^B.
\]
If $A=X$ is classical and $\cP^A$ is its classical eigenbasis measurement, the resulting state is CQ in the same basis.
\end{lemma}
The proof of Lemma \ref{Lem:qustwithclmarg} is given in Section \ref{Sec:Lem:qustwithclmarg-proof} below.

\medskip

\noindent \textbf{Part $(i)-$Converse:} 
The achievability of \eqref{eq:zeroratesteinexp} is already shown in \eqref{eq:achievabilityzerorate} above. So, we will  prove the converse part, i.e.,   $\theta_{\mathsf{C}}(\epsilon, \rho_{AB}, \tilde \rho_{AB}) \leq \theta^{\star}_{\mathsf{PLO}\text{-}\rho}(\rho_{AB},\tilde \rho_{AB})$ for all $\epsilon \in [0,1)$ when the support and product-marginal eigenbasis commutativity conditions of Theorem~\ref{Thm:Steinsexp-classical-zerorate}(i) holds. Recall the definition of  $\theta^{\star}_{\mathsf{BPLO}\text{-}\rho}(\rho_{AB},\tilde \rho_{AB})$ given in \eqref{eq:limitexp-projlocstein}. It suffices to show  that
\begin{subequations} \label{eq:compbpoandpo}
\begin{align}
    \theta_{\mathsf{C}}(\epsilon, \rho_{AB}, \tilde \rho_{AB})  
   & \leq \theta^{\star}_{\mathsf{BPLO}\text{-}\rho}(\rho_{AB},\tilde \rho_{AB}),~ \forall ~\epsilon \in [0,1), \label{eq:converse-zerorate}
\end{align}
\text{and} 
  \begin{align}
\theta^{\star}_{\mathsf{BPLO}\text{-}\rho}(\rho_{AB},\tilde \rho_{AB}) \leq \theta^{\star}_{\mathsf{PLO}\text{-}\rho}(\rho_{AB},\tilde \rho_{AB}).  \label{eq:dpilogsumineq}
  \end{align}
\end{subequations}
  Equation \eqref{eq:converse-zerorate} is  a  consequence of the non-asymptotic strong converse lemma stated below, which we prove in Section \ref{Sec:Lem:strong-converse-proof} using the bipartite generalization (Lemma \ref{lem:quant-blowup-bipartite}) of the classical blowing-up lemma,  referred to henceforth as the \textit{commuting product marginal eigenbasis blowing-up lemma} (CPME-BL). 
\begin{lemma}[Non-asymptotic strong converse via CPME-BL] \label{Lem:strong-converse}
   Let  $(r_n)_{n \in \NN}$ be a sequence such that $1-2e^{-2r_n^2}>0$ for every $n$ under consideration (equivalently, $r_n>\sqrt{\log 2/2}$), and let $\rho_{AB}$, $\tilde \rho_{AB}$ be such that the support and product-marginal eigenbasis commutativity conditions in Theorem \ref{Thm:Steinsexp-classical-zerorate}$(i)$ hold. Then, for any $\cF_n^{A^n \to W_n}=\cM_n^{A^n \to X_n}(\cdot)$  (with $W_n=X_n$,  $|W_n|=\abs{\cX_n}>1$ and $\cM_n^{A^n \to X_n}(\cdot)$ being a measurement channel),  $\cG_n,\cT_n$ such that $ \alpha_n(\cF_n,\cG_n,\cT_n) \leq \epsilon < 1$, we have
   \begin{align}
   -\frac{1}{n}\log \beta_n(\cF_n,\cG_n,\cT_n) &\leq  \sup_{\cM_n \in \mathsf{BPLO}^{\rho}_n}\inf_{\substack{\hat \rho^n_{AB}   \in \mspace{2 mu}\cD_n(\rho_{AB})}}\frac{\qrel{\cM_n(\hat \rho^n_{AB})}{\cM_n(\tilde  \rho_{AB}^{\otimes n})}}{n(1-2e^{-2r_n^2})}+  \frac{2}{n}\log \gamma_n(\epsilon_n,r_n,\tilde \rho_{AB}) \notag \\
   & \qquad \qquad \qquad \qquad \qquad \qquad \qquad \qquad \qquad \qquad +\frac{1}{n(1-2e^{-2r_n^2})}, \label{eq:type2err-strongconv}
\end{align}
where $\gamma_n(\epsilon_n,r_n,\tilde \rho_{AB})$ is as defined in \eqref{eq:blowupfactor}, 
\begin{align}
\epsilon_n=\frac{1-\epsilon}{|W_n|} \quad \mbox{ and } \quad   l_n (\epsilon_n,r_n)&=\left\lceil\sqrt{0.5 n\big(\log 2- \log \epsilon_n\big)}+r_n\sqrt{n}\right\rceil. \label{eq:parambup}
\end{align} 
\end{lemma}

\medskip

To prove \eqref{eq:converse-zerorate} from \eqref{eq:type2err-strongconv}, 
it will suffice to show that $\log \gamma_n(\epsilon_n,r_n)/n \rightarrow 0$  for $\epsilon_n=(1-\epsilon)/|W_n|$ and some $r_n \rightarrow \infty$ when the support and product-marginal eigenbasis commutativity conditions hold. Let  $r_n=n^{1/3}$. Note that $\rho_A \otimes \rho_B \ll \tilde \rho_{AB}$ implies that $\bar{\mu}_{\min}(\tilde \rho_{AB})>0$.   Using $\binom{n}{l} \leq (ne/l)^l$, we have  
\begin{align}
  \gamma_{n}(\epsilon_n,r_n,\sigma)&:=\frac{2(d_A \vee d_B)^{l_n(\epsilon_n,r_n)} \sum_{l=0}^{l_n(\epsilon_n,r_n)}\binom{n}{l}}{\epsilon_n\big(\bar{\mu}_{\min}(\tilde \rho_{AB})\big)^{l_n(\epsilon_n,r_n)}} \leq  \frac{2(d_A \vee d_B)^{l_n(\epsilon_n,r_n)} (l_n(\epsilon_n,r_n)+1)(ne)^{l_n(\epsilon_n,r_n)} }{ \epsilon_n\big(\bar{\mu}_{\min}(\tilde \rho_{AB}) l_n(\epsilon_n,r_n)\big)^{l_n(\epsilon_n,r_n)} }. \notag
\end{align}
The claim then follows by noting that for the given choice of $r_n$, $l_n(\epsilon_n,r_n)=o(n)$ since $\log |W_n|=o(n)$ due to \eqref{eq:zeroratequant}. This implies that 
\begin{align}
 0 \leq   \frac{  \log \gamma_n(\epsilon_n,r_n)}{n} & \leq  \frac{l_n(\epsilon_n,r_n) \log (d_A \vee  d_B)+\log \left(\frac{2}{\epsilon_n}\right)}{n}+ \frac{\log (l_n(\epsilon_n,r_n)+1) }{n}+\frac{l_n(\epsilon_n,r_n)}{n}\log \left(\frac{e}{\bar{\mu}_{\min}(\tilde \rho_{AB})}\right) \notag \\
 & \qquad - \frac{l_n(\epsilon_n,r_n)}{n}\log \left(\frac{l_n(\epsilon_n,r_n)}{n}\right) \notag \\
 &\rightarrow 0. \notag 
\end{align}
Using these, \eqref{eq:converse-zerorate} follows from \eqref{eq:type2err-strongconv} by maximizing the LHS over $\cF_n,\cG_n,\cT_n$ and taking limit superior over $n$. 

\medskip

To complete the proof of \eqref{eq:zeroratesteinexp}, it remains to show \eqref{eq:dpilogsumineq}. For this, consider 
 \begin{align}
  \cP_n=\cP_n^{A^n} \otimes \cP_n^{B^n}=\big\{P_x^{A^n}\big\}_{x \in [1:d_A^n]} \otimes \big\{P_{y}^{B^n}\big\}_{y \in [1:d_B^n]}  \in \mathsf{PLO}^{\rho}_n, \notag
 \end{align}
 and  for some $\hat{\cX}_n \subseteq [1:d_A^n]$ and $\hat{\cY}_n \subseteq [1:d_B^n]$, 
\begin{align}
    \cM_n=\left\{\sum_{x \in  \hat{\cX}_n} P_{x}^{A^n},I_{A^n}-\sum_{x \in  \hat{\cX}_n }P_{x}^{A^n}\right\} \otimes \left\{\sum_{y \in \hat{\cY}_n} P_{y}^{B^n},I_{B^n}-\sum_{y \in \hat{\cY}_n }P_{y}^{B^n}\right\} \in \mathsf{BPLO}^{\rho}_n. \notag
\end{align}
For any $\hat \rho^n_{AB}$,  we have by an application of the log-sum inequality \cite[Theorem 2.7.1]{CoverThomas} or the data-processing inequality for quantum relative entropy that
\begin{align}
\qrel{\cM_n(\hat \rho^n_{AB})}{\cM_n(\tilde \rho_{AB}^{\otimes n})} \leq \qrel{\cP_n(\hat \rho^n_{AB})}{\cP_n(\tilde \rho_{AB}^{\otimes n})}. \notag
\end{align}
Taking infimum over
$\hat \rho^n_{AB} \in \cD_n(\rho_{AB})$ and supremum with respect to $\cM_n \in \mathsf{BPLO}^{\rho}_n$ yields 
\begin{align}
\sup_{ \cM_n \in \mathsf{BPLO}^{\rho}_n} \inf_{\hat \rho^n_{AB} \in \cD_n(\rho_{AB})}  \qrel{\cM_n(\hat \rho^n_{AB})}{\cM_n(\tilde \rho_{AB}^{\otimes n})} 
 &\leq \sup_{\cP_n \in \mathsf{PLO}^{\rho}_n} \inf_{\hat \rho^n_{AB} \in \cD_n(\rho_{AB})} \qrel{\cP_n(\hat \rho^n_{AB})}{\cP_n(\tilde \rho_{AB}^{\otimes n})}. \notag
\end{align}
Taking limit superior  over $n$, we obtain \eqref{eq:dpilogsumineq}. Note that \eqref{eq:compbpoandpo} along with \eqref{eq:achievabilityzerorate} implies that the limit in \eqref{eq:limitexp-projlocstein} exists.

\medskip

\noindent \textbf{Part $(ii)-$Achievability:} To show the lower bound in \eqref{eq:zerorateweakconv}, we will use \eqref{eq:multilett-char}. By restricting to encoders  $\cF_n^{A^n \to X_n}=\cM_n^{A^n \to X_n}$ and   $\cG_n^{B^n \to Y_n}=\cM_n^{B^n \to Y_n}$ for measurement channels  $\cM_n^{A^n \to X_n}$ and $\cM_n^{B^n \to Y_n}$ satisfying the zero-rate communication constraint, we obtain
\begin{align}
   \lim_{\epsilon \downarrow 0^+} \theta_{\mathsf{C}}(\epsilon, \rho_{AB}, \tilde \rho_{AB}) &\geq  \limsup_{n \rightarrow \infty} \sup_{\cM_n \in  \mathsf{LO-ZR_n}} \frac{\qrel{\cM_n( \rho_{AB}^{\otimes n})}{\cM_n(\tilde \rho_{AB}^{\otimes n})}}{n}, \label{eq:achiev-zr} 
\end{align}
where $\mathsf{LO-ZR_n}$ is the set of all local POVMs  $\cM_n=\cM_n^{A^n \to X_n} \otimes \cM_n^{B^n \to Y_n}=\{M_{x}^{A^n}\}_{x \in \cX_n} \otimes \{M_{y}^{B^n}\}_{y \in \cY_n}$. 

Now, consider the binary outcome POVM  $\check{\cM}_n=\{\check M_n, I-\check M_n\}$ with $\check M_n=\sum_{x,y \in \bar{\cX}_n \times \bar{\cY}_n} \ket{x}\bra{x} \otimes \ket{y}\bra{y}$ for some $\bar \cX_n \subseteq \cX_n$ and $\bar \cY_n \subseteq \cY_n$. By the data-processing inequality applied to the quantum channel induced by $\check{\cM}_n$, we obtain
\begin{align}
 &\qrel{\cM_n( \rho_{AB}^{\otimes n})}{\cM_n(\tilde \rho_{AB}^{\otimes n})} \notag \\
 & \geq  \tr{\check{M}_n \cM_n\big(\rho_{AB}^{\otimes n}\big) } \log \left(\frac{\tr{\check{M}_n \cM_n\big(\rho_{AB}^{\otimes n}\big)}}{\tr{\check{M}_n \cM_n\big(\tilde \rho_{AB}^{\otimes n}\big)}}\right)+\big(1-\tr{\check{M}_n \cM_n\big(\rho_{AB}^{\otimes n}\big)}\big) \log \left(\frac{1-\tr{\check{M}_n \cM_n\big(\rho_{AB}^{\otimes n}\big)}}{1-\tr{\check{M}_n \cM_n\big(\tilde \rho_{AB}^{\otimes n}\big)}}\right) \notag \\
 & =\tr{\cM_n^{\dag}\big(\check{M}_n\big) \rho_{AB}^{\otimes n} } \log \left(\frac{\tr{\cM_n^{\dag}\big(\check{M}_n\big) \rho_{AB}^{\otimes n}}}{\tr{\cM_n^{\dag}\big(\check{M}_n\big) \tilde \rho_{AB}^{\otimes n}}}\right) +\big(1-\tr{\cM_n^{\dag}\big(\check{M}_n\big) \rho_{AB}^{\otimes n}\big)} \log \left(\frac{1-\tr{\cM_n^{\dag}\big(\check{M}_n \big)\rho_{AB}^{\otimes n}}}{1-\tr{\cM_n^{\dag}\big(\check{M}_n \big)\tilde \rho_{AB}^{\otimes n}}}\right), \notag 
\end{align}
where $\cM_n^{\dag}$ is the adjoint map of $\cM_n$. 
Note that
\begin{align}
  &\cM_n^{\dag}\big(\check{M}_n\big)\notag \\
  &=\sum_{x,x',y,y'} \left(\sqrt{M_x^{A^n}} \ket{x'}\bra{x} \otimes \sqrt{M_y^{B^n}} \ket{y'}\bra{y}\right) \left(\sum_{(\bar x,\bar y) \in \bar{\cX}_n \times \bar{\cY}_n}\ket{\bar x}\bra{\bar x} \otimes  \ket{\bar y}\bra{\bar y}\right) \left(\ket{x}\bra{x'} \sqrt{M_x^{A^n}} \otimes \ket{y}\bra{y'} \sqrt{M_y^{B^n}} \right)\notag \\
  &=\sum_{(\bar x,\bar y) \in \bar{\cX}_n \times \bar{\cY}_n} M_x^{A^n}  \otimes M_y^{B^n}. \notag 
\end{align}
Substituting this in the equation above, and taking supremum over $\cM_n \in \mathsf{LO-ZR_n}$ and $\check M_n$, we obtain the desired lower bound  from \eqref{eq:achiev-zr} using the definition of $\theta^{\star}_{\mathsf{BSEP-ZR}}(\rho_{AB},\tilde \rho_{AB})$ given in \eqref{eq:limitexp-sepzrexp}.

\medskip

\noindent \textbf{Part $(ii)-$Converse:} Next, we prove the converse part of \eqref{eq:zerorateweakconv}. Recall that $ \theta_{\mathsf{C}}(\epsilon, \rho_{AB}, \tilde \rho_{AB}) $ denotes the Stein's exponent, where one of the encoders (Alice here) communicates classical information at zero-rate to Charlie. Hence, Charlie has access to $X_n,V_n$, where $X_n $ is the output of the measurement channel $\cF_n^{A^n \to X_n}=\cM_n^{A^n \to X_n}$. We can assume without loss of generality (for the converse) that Charlie has direct  access to $B^n$, i.e., $V_n=B^n$.   
Since the joint state on  $(X_n,B^n)$ is a CQ state, the most general test  $\cT_n=\{T_n,I-T_n\}$ at Charlie can be written in the form $T_n=\sum_{x \in  \cX_n} \ket{x}\bra{x} \otimes M_x^{B^n}$ for some set $\cX_n$ such that  $0 \leq  M_x^{B^n} \leq I_{B^n}$ for all $x \in \cX_n $.

Fix some sequence $(\bar \epsilon_n)_{n \in \NN}$ such that $\bar \epsilon_n \downarrow 0^+$,  and $(\cF_n,\cG_n,\cT_n)_{n \in \NN}$ be such that $\alpha_n(\cF_n,\cG_n,\cT_n) \leq \bar \epsilon_n$. We have
\begin{align}
    \tr{ (\cM_n \otimes \cI_n) (\rho_{AB}^{\otimes n})~ T_n} =  1-\alpha_n(\cF_n,\cG_n,\cT_n). \notag
\end{align}
 This means 
\begin{align}
    \tr{\rho_{AB}^{\otimes n} \sum_{x,x'} \left(\sqrt{M_x^{A^n}} \ket{x'}\bra{x} \otimes I\right) \big(\sum_{\bar x \in \cX_n}\ket{\bar x}\bra{\bar x} \otimes  M_{\bar x}^{B^n}\big) \left(\ket{x}\bra{x'} \sqrt{M_x^{A^n}} \otimes I  \right)} = 1-\alpha_n(\cF_n,\cG_n,\cT_n), \notag
\end{align}
which further leads to 
\begin{align}
    \tr{\rho_{AB}^{\otimes n} \sum_{ x \in \cX_n}\big( M_x^{A^n} \otimes M_x^{B^n}  \big)}  = 1-\alpha_n(\cF_n,\cG_n,\cT_n). \label{eq:t1errcl} 
\end{align}
Now, with $\widebar{M}_n=\sum_{x \in \cX_n} M_x^{A^n} \otimes M_x^{B^n}$, consider  $\widebar{\cM}_n= \left\{\widebar{M}_n,I_{A^nB^n}-\widebar{M}_n\right\} \in \mathsf{BSEP-ZR_n}$.
Then,  we have
\begin{align}
  &  \qrel{\widebar{\cM}_n(\rho_{AB}^{\otimes n})}{\widebar{\cM}_n(\tilde \rho_{AB}^{\otimes n})} \notag \\
  &= \tr{\widebar{M}_n \rho_{AB}^{\otimes n} } \log \left(\frac{\tr{\widebar{M}_n \rho_{AB}^{\otimes n}}}{\tr{\widebar{M}_n \tilde \rho_{AB}^{\otimes n}}}\right)+\big(1-\tr{\widebar{M}_n \rho_{AB}^{\otimes n} }\big) \log \left(\frac{1-\tr{\widebar{M}_n \rho_{AB}^{\otimes n}}}{1-\tr{\widebar{M}_n \tilde \rho_{AB}^{\otimes n}}}\right) \notag \\
    &=\big(1-\alpha_n(\cF_n,\cG_n,\cT_n) \big) \log \left(\frac{1-\alpha_n(\cF_n,\cG_n,\cT_n)}{\beta_n(\cF_n,\cG_n,\cT_n)}\right)+\alpha_n(\cF_n,\cG_n,\cT_n) \log \left(\frac{\alpha_n(\cF_n,\cG_n,\cT_n)}{1-\beta_n(\cF_n,\cG_n,\cT_n)}\right) \notag \\
    & \geq  -h_b \left(\alpha_n(\cF_n,\cG_n,\cT_n)\right)-\big(1-\alpha_n(\cF_n,\cG_n,\cT_n) \big) \log \left(\beta_n(\cF_n,\cG_n,\cT_n)\right), \notag
\end{align}
where $h_b(\cdot) $ is the classical binary entropy defined as $h_{b}(p):=-p \log p-(1-p) \log (1-p)$. 
Hence
\begin{align}
    -\frac{1}{n } \log \big(\beta_n(\cF_n,\cG_n,\cT_n)\big) \leq \frac{1}{\big(1-\alpha_n(\cF_n,\cG_n,\cT_n) \big)} \frac{\qrel{\widebar{\cM}_n(\rho_{AB}^{\otimes n})}{\widebar{\cM}_n(\tilde \rho_{AB}^{\otimes n})}}{n}+ \frac{h_b \left(\alpha_n(\cF_n,\cG_n,\cT_n)\right)}{n\big(1-\alpha_n(\cF_n,\cG_n,\cT_n) \big)}. \notag
\end{align}
Minimizing  $\beta_n(\cF_n,\cG_n,\cT_n)$ over  all $(\cF_n,\cG_n,\cT_n)$ such that $ \alpha_n(\cF_n,\cG_n,\cT_n) \leq \bar 
 \epsilon_n$ and noting that $0 \leq h_b(\alpha) \leq 1$ for all $0 \leq \alpha \leq 1$, we have
\begin{align}
    -\frac{1}{n } \log \left(\bar \beta_n(\epsilon_n)\right) \leq \frac{1}{\big(1-\bar \epsilon_n \big)} \sup_{\widebar{\cM}_n \in  \mathsf{BSEP-ZR_n}} \frac{\qrel{\widebar{\cM}_n(\rho_{AB}^{\otimes n})}{\widebar{\cM}_n(\tilde \rho_{AB}^{\otimes n})}}{n}+ \frac{1}{n\big(1-\bar \epsilon_n \big)}. \notag 
\end{align}
Taking limit superior with respect to $n$ yields
\begin{align}
    \limsup_{n \rightarrow \infty} -\frac{1}{n } \log \left(\bar \beta_n(\bar \epsilon_n)\right) \leq \theta^{\star}_{\mathsf{BSEP-ZR}}(\rho_{AB},\tilde \rho_{AB}).  \notag 
\end{align}
Since $\bar  \epsilon_n \downarrow 0^+$ is arbitrary, this proves the upper bound in \eqref{eq:zerorateweakconv}. 
\subsection{Proof of Corollary \ref{Cor:zeroSteinexp}}
Recalling the definition of $\theta^{\star}_{\mathsf{PLO}\text{-}{\rho}}(\rho_{AB},\tilde \rho_{AB})$ in \eqref{eq:optimalmultlettexp},   observe that for  any   $\rho_{AB}$ and $\tilde \rho_{AB}$ such that $\tilde \rho_{A}=\rho_{A}$, $\tilde \rho_{B}=\rho_{B}$ and  $\rho_{A} \otimes \rho_{B} \ll \tilde \rho_{AB}$, we have
\begin{align}
 0 \leq   \theta^{\star}_{\mathsf{PLO}\text{-}{\rho}}(\rho_{AB},\tilde \rho_{AB}) &=\lim_{n \rightarrow \infty}\max_{\cM_n \in  \mathsf{PLO}_n^{\rho}} \min_{\hat \rho^n_{AB} \in \cD_n(\rho_{AB})} \frac{\qrel{\cM_n(\hat \rho^n_{AB})}{\cM_n(\tilde \rho_{AB}^{\otimes n})}}{n} \notag \\
 & \leq \lim_{n \rightarrow \infty}\max_{\cM_n \in  \mathsf{PLO}_n^{\rho}}  \frac{\qrel{\cM_n(\tilde  \rho_{AB}^{\otimes n})}{\cM_n(\tilde \rho_{AB}^{\otimes n})}}{n} \notag \\
 &=0, \notag
\end{align}
where the inequality follows since $\tilde \rho_{AB}^{\otimes n} \in \cD_n(\rho_{AB})$ for every $\cM_n$ and $n \in \NN$. Hence, $\theta^{\star}_{\mathsf{PLO}\text{-}{\rho}}(\rho_{AB},\tilde \rho_{AB})=0$. The claim then follows from \eqref{eq:zeroratesteinexp} under the support and product-marginal eigenbasis commutativity conditions of Theorem~\ref{Thm:Steinsexp-classical-zerorate}(i).  When $\rho_{AB}= \tilde \rho_{AB}$, we have 
\begin{align}
 0 \leq  \theta_{\mathsf{C}}(\epsilon, \rho_{AB}, \tilde \rho_{AB}) \leq \qrel{\rho_{AB}}{\tilde \rho_{AB}}=0, \notag
\end{align}
where the inequality is because  $\qrel{\rho_{AB}}{\tilde \rho_{AB}}$, which is the Stein's exponent in the centralized setting, is an upper bound on $\theta(\epsilon, \rho_{AB}, \tilde \rho_{AB})$ for $0 \leq \epsilon<1$.

\subsection{Proof of Corollary \ref{Cor:Steinexpcqstates}}\label{Sec:Cor:Steinexpcqstates-proof}
Consider the local projective measurement channel
$\cM_n=\cP_n^{\star A^n}\otimes\cP_n^{B^n}\in\mathsf{PLO}_n$, where $\cP_n^{B^n}=\{P_{y_n}^{B^n}=\ket{y_n}\bra{y_n}\}_{y_n\in\cY_n}$ is a rank-one projective measurement commuting with $\rho_B^{\otimes n}$ and $\cP_n^{\star A^n}$ is the measurement in the classical basis $\{\ket{x^n}\bra{x^n}\}_{x^n\in\cX^n}$. Using the same argument as in the proof of \eqref{eq:achievabilityzerorate}, with Lemma~\ref{Lem:qustwithclmarg}(ii), gives for every $\epsilon\in(0,1]$
\begin{align}
\theta_{\mathsf C}(\epsilon,\rho_{XB},\tilde\rho_{XB})
&\geq \lim_{n\to\infty}\max_{\cP_n^{B^n}\in\mathsf P^{\rho_B}_n(B^n)}
\min_{\hat\rho^n_{XB}\in\tilde{\cD}_n(\rho_{XB})}
\frac{\qrel{(I_{X^n}\otimes\cP_n^{B^n})(\hat\rho^n_{XB})}{(I_{X^n}\otimes\cP_n^{B^n})(\tilde\rho_{XB}^{\otimes n})}}{n}\notag\\
&=\theta^\star_{\mathsf{CQ}}(\rho_{XB},\tilde\rho_{XB}).\notag
\end{align}
Moreover, every binary projective measurement in $\mathsf{BP}_n^{\rho_B}(B^n)$ is a coarse graining of a rank-one projective measurement in $\mathsf P^{\rho_B}_n(B^n)$.  The log-sum inequality therefore yields
\begin{align}
\theta^\star_{\mathsf{CQ}}(\rho_{XB},\tilde\rho_{XB})\geq
\theta^\star_{\mathsf{CQ-BP}}(\rho_{XB},\tilde\rho_{XB}),\notag
\end{align}
which proves \eqref{eq:SteinexpCQchar}.  The proof of the converse bound $\theta_{\mathsf C}(\epsilon,\rho_{XB},\tilde\rho_{XB}) \leq 
\theta^\star_{\mathsf{CQ-BP}}(\rho_{XB},\tilde\rho_{XB})$ leading to equality in \eqref{eq:SteinexpCQchar} follows similar to \eqref{eq:converse-zerorate} using Lemma \ref{Lem:strong-converse} under the support and commutativity conditions.

\subsection{Proof of Proposition \ref{Prop:counterexcqstate}}\label{Sec:Prop:counterexcqstate-proof}
We prove the desired upper bound directly as the relevant bound in Corollary~\ref{Cor:Steinexpcqstates}, which relies on  CPME-BL, does not apply in general when $\tilde \rho_0$ and $\tilde \rho_1$ are  density operators of full rank (the commutativity condition may not hold).  
Without loss of generality, we may assume that  $B^n$ system is available at Charlie for the converse bound. Since the $X^n$ system is classical, Alice's classical encoder can be written as a conditional distribution $P_{W_n|X^n}$ with $\log|W_n|=o(n)$. The most general test at  Charlie can be written as $\cT_n=\{T_n,I-T_n\}$, where  $T_n=\sum_{w_n \in  \cW_n} \ket{w_n}\bra{w_n} \otimes M_{w_n}^{B^n}$ with $0\leq M_{w_n}^{B^n}\leq I_{B^n}$.  Under the null hypothesis, the joint state is of the form $\rho_{X}^{\otimes n} \otimes \rho_B^{\otimes n}=\rho_{X}^{\otimes n} \otimes \tau^{\otimes n}$, where $\rho_X$ is a classical state  uniformly distributed with support $\{0,1\}$ and $\tau$ is an arbitrary state.  Denoting the   test by $(\cF_n,\cG_n,\cT_n)_{n \in \NN}$ and using the shorthands $\alpha_n=\alpha_n(\cF_n,\cG_n,\cT_n)$ and $\beta_n=\beta_n(\cF_n,\cG_n,\cT_n)$ for the resulting type I and type II error probabilities, we have
\begin{align}
1-\alpha_n
&=\sum_{w_n\in\cW_n} r_{w_n}\,m_{w_n}\geq 1-\epsilon, \notag  \\
\text{where} \quad r_{w_n}&:=2^{-n}\sum_{x^n \in \cX^n}P_{W_n|X^n}(w_n|x^n) \quad \text{and} \quad 
m_{w_n}:=\tr{\tau^{\otimes n}M_{w_n}^{B^n}}. \notag
\end{align}
Therefore, for some $w_n^\star\in\cW_n$,
\begin{align}
r_{w_n^\star}m_{w_n^\star}\geq \delta_n,
\qquad \delta_n:=\frac{1-\epsilon}{|W_n|}.\notag
\end{align}
Note that $r_{w_n^\star}\geq\delta_n$, $m_{w_n^\star}\geq\delta_n$, and $-\log\delta_n=o(n)$.  Set $r:=r_{w_n^\star}$, $M_n:=M_{w_n^\star}^{B^n}$ and
\begin{align}
\cC_n:=\left\{x^n\in\{0,1\}^n:\ P_{W_n|X^n}(w_n^\star|x^n)\geq r/2\right\}. \notag
\end{align}
Since $2^{-n}\sum_{x^n}P_{W_n|X^n}(w_n^\star|x^n)=r$ and the summands are at most one,
\begin{align}
\frac{|\cC_n|}{2^n}\geq \frac{r}{2-r}\geq \frac r2\geq\frac{\delta_n}{2}.\label{eq:prop8-cellsize}
\end{align}
Moreover, the type-II error probability satisfies
\begin{align}
\beta_n
&\geq \sum_{x^n \in \cX^n}\tilde p_{X^n}(x^n)P_{W_n|X^n}(w_n^\star|x^n)
\tr{M_n\tilde\rho_{x^n}}\geq \frac{\delta_n}{2}\sum_{x^n\in\cC_n}
\tilde p_{X^n}(x^n)\tr{M_n\tilde\rho_{x^n}},\label{eq:prop8-beta-cell}
\end{align}
where $\tilde\rho_{x^n}:=\bigotimes_{i=1}^n\tilde\rho_{x_i}$.

Let $\bar x^n$ denote the bitwise complement of $x^n$.  By the classical blowing-up lemma \cite{Ahlswede-Gacs-Korner-1976,Csiszar-Korner,Marton-1986}, \eqref{eq:prop8-cellsize} and $-\log\delta_n=o(n)$ imply that there exists a sequence $\ell_n=o(n)$ such that the Hamming $\ell_n$-neighborhood $\cC_n^{+\ell_n}$ satisfies
\begin{align}
2^{-n}|\cC_n^{+\ell_n}|\geq 1-\frac{\delta_n}{4}.\label{eq:prop8-blowup}
\end{align}
Since $|\overline{\cC}_n|=|\cC_n|$, it follows from \eqref{eq:prop8-cellsize}--\eqref{eq:prop8-blowup} that
\begin{align}
|\overline{\cC}_n\cap\cC_n^{+\ell_n}|\geq \frac{\delta_n}{4}2^n.\label{eq:prop8-intersection}
\end{align}
Consequently, there is a set $\tilde \cC_n \subseteq\cC_n$ with $|\tilde \cC_n|\geq \delta_n2^{n-2}$ and a map $y^n:\tilde \cC_n\to\cC_n$ such that
\begin{align}
d_{\mathsf{H}}\big(y^n(x^n),\bar x^n\big)\leq\ell_n,\qquad x^n\in \tilde \cC_n.\label{eq:prop8-pairing}
\end{align}
Each $y_n\in\cC_n$ has at most
\begin{align}
Z_n(\ell_n):=\sum_{j=0}^{\ell_n}{n\choose j}=e^{o(n)}\label{eq:prop8-volume}
\end{align}
preimages under this map.

 We will use a standard inequality for the Kubo--Ando geometric mean
\begin{align}
\tr{L\,\omega(A,B)}\leq \sqrt{\tr{LA}\tr{LB}},\qquad L,A,B\geq0,\label{eq:prop8-geomean-functional}
\end{align}
which follows by applying \cite[Theorem 3]{Ando-concavity} to  the positive functional $A\mapsto\tr{LA}$. Since $\tilde\rho_0,\tilde\rho_1$ are full rank and $0<\tilde p_X(0),\tilde p_X(1)<1$, there is a constant $c\geq1$ such that for $x\in \{0,1\}$,
\begin{align}
\tilde p_X(x)\, \tilde\rho_x \geq c^{-1}\sqrt{\tilde p_X(0)\,\tilde p_X(1)}\,\omega\big(\tilde \rho_0,\tilde \rho_1\big).\label{eq:prop8-local-domination}
\end{align}
For $x^n\in \tilde \cC_n$, all but at most $\ell_n$ coordinates of $y^n(x^n)$ are complementary to the corresponding coordinates of $x^n$.  
Using symmetry and tensorization of the geometric mean together with \eqref{eq:prop8-local-domination}, we obtain
\begin{align}
\sqrt{\tilde p_{X^n}(x^n)\tilde p_{X^n}(y^n(x^n))}
\,\omega\big(\tilde\rho_{x^n},\tilde\rho_{y^n(x^n)}\big)&=
\,\bigotimes_{i=1}^n\, \sqrt{\tilde p_{X_i}(x_i)\tilde p_{X_i}(y_i(x^n))}\,\omega\big(\tilde\rho_{x_i},\tilde\rho_{y_i(x^n)}\big) \notag \\
& \geq c^{-\ell_n}\big(\tilde p_{X}(0)\tilde p_{X}(1))^{\frac n2}\,\omega\big(\tilde\rho_0,\tilde\rho_1\big)^{\otimes n}.\label{eq:prop8-pair-domination}
\end{align}
Define
\begin{align}
a_{x^n}:=\tilde p_{X^n}(x^n)\tr{M_n\tilde\rho_{x^n}},\qquad x^n\in\cC_n. \notag 
\end{align}
Equations \eqref{eq:prop8-geomean-functional}--\eqref{eq:prop8-pair-domination} and the arithmetic--geometric mean inequality imply, for every $x^n\in \tilde \cC_n$,
\begin{align}
a_{x^n}+a_{y^n(x^n)}
\geq 2\,c^{-\ell_n}\big(\tilde p_{X}(0)\tilde p_{X}(1))^{\frac n2}\tr{M_n\omega\big(\tilde\rho_0,\tilde\rho_1\big)^{\otimes n}}.\notag 
\end{align}
Summing over $x^n\in \tilde \cC_n$ and using \eqref{eq:prop8-volume},
\begin{align}
\big(1+Z_n(\ell_n)\big)\sum_{x^n\in \cC_n}a_{x^n}
&\geq \frac{\delta_n}{2}2^n c^{-\ell_n}\big(\tilde p_{X}(0)\tilde p_{X}(1))^{\frac n2} \tr{M_n\omega\big(\tilde\rho_0,\tilde\rho_1\big)^{\otimes n}}.\label{eq:prop8-summedpairs}
\end{align}
Combining \eqref{eq:prop8-beta-cell} and \eqref{eq:prop8-summedpairs} yields
\begin{align}
\beta_n
\geq \frac{\delta_n^2}{4\big(1+Z_n(\ell_n)\big)}c^{-\ell_n}
\big(2\sqrt{\tilde p_{X}(0)\tilde p_{X}(1)}\big)^n\tr{M_n\omega\big(\tilde\rho_0,\tilde\rho_1\big)^{\otimes n}}.\label{eq:prop8-beta-lower}
\end{align}
We will next lower bound the last trace in the RHS of the above equation in terms of sandwiched R\'enyi divergence. For $\alpha>1$, the sandwiched R\'enyi divergence
\[
D_\alpha^\ast(\rho\|\sigma)
:=\frac{1}{\alpha-1}\log\tr{\left(
\sigma^{\frac{1-\alpha}{2\alpha}}\rho
\sigma^{\frac{1-\alpha}{2\alpha}}\right)^\alpha}.
\]
For any $0\leq L_n\leq I$, set
$s_n:=\tr{\rho^{\otimes n}L_n}$ and
$q_n:=\tr{\sigma^{\otimes n}L_n}$. By data processing inequality for
$D_\alpha^\ast$ under the binary measurement $\{L_n,I-L_n\}$ and by
retaining only the first term in the resulting binary R\'enyi divergence,
\begin{align}
nD_\alpha^\ast(\rho\|\sigma)
&\geq
\frac{1}{\alpha-1}
\log\!\left(s_n^\alpha q_n^{1-\alpha}
 +(1-s_n)^\alpha(1-q_n)^{1-\alpha}\right)\geq
\frac{\alpha}{\alpha-1}\log s_n-\log q_n .
\end{align}
Consequently,
\begin{align}
-\frac1n\log q_n
\leq D_\alpha^\ast(\rho\|\sigma)
+\frac{\alpha}{\alpha-1}\frac{-\log s_n}{n}.
\label{eq:prop8-one-system-renyi}
\end{align}
Apply this with $\rho=\tau$,
$\sigma=\omega(\tilde\rho_0,\tilde\rho_1)$,
$L_n=M_n$, and $s_n=m_{w_n^\star}\geq\delta_n$.
Since $\tilde\rho_0$ and $\tilde\rho_1$ are full rank, their geometric mean
$\omega(\tilde\rho_0,\tilde\rho_1)$ is full rank. Put
$a_n:=-\log\delta_n=o(n)$ and choose
\[
\alpha_n:=1+\sqrt{\frac{a_n+1}{n}}.
\]
Then $\alpha_n\downarrow1$,
$D_{\alpha_n}^\ast(\tau\|\omega(\tilde\rho_0,\tilde\rho_1))
\to D(\tau\|\omega(\tilde\rho_0,\tilde\rho_1))$, and
\[
\frac{\alpha_n}{\alpha_n-1}\frac{a_n}{n}
=
\frac{a_n}{n}
+\frac{a_n}{\sqrt{n(a_n+1)}}\longrightarrow0.
\]
Thus
\begin{align}
-\frac1n\log\tr{M_n\omega\big(\tilde\rho_0,\tilde\rho_1\big)^{\otimes n}}
\leq \qrel{\tau}{\omega\big(\tilde\rho_0,\tilde\rho_1\big)}+o(1).
\label{eq:prop8-omega-converse}
\end{align}
Noting that $\qrel{p_X}{\tilde p_X}=-\log\big(2\sqrt{\tilde p_{X}(0)\tilde p_{X}(1)}\big)$,  $-\log\delta_n=o(n)$, $\ell_n=o(n)$, and $\log Z_n(\ell_n)=o(n)$,  \eqref{eq:prop8-beta-lower}--\eqref{eq:prop8-omega-converse} yields
\begin{align}
\theta_{\mathsf C}(\epsilon,\rho_{XB},\tilde\rho_{XB})
\leq \qrel{p_X}{\tilde p_X}+\qrel{\tau}{\omega(\tilde\rho_0,\tilde\rho_1)}.\notag 
\end{align}
It remains to show the final equality in \eqref{eq:steinexpgapspec} when $\tau=\psi$.  To this end, we obtain an expression for $\theta^{\star}_{\mathsf{SL}}(\rho_{XB},\tilde \rho_{XB})$. 
Since  $\rho_0=\rho_1=\psi$ for some pure state $\psi=\ket{\psi}\bra{\psi}$ by assumption, 
\begin{align} \rho_{XB}=\left(\sum_{x=0}^1 \frac 12 \ket{x}\bra{x}\right) \otimes \ket{\psi}\bra{\psi}. \notag 
\end{align}
Consider an arbitrary 
\begin{align}
    \hat \rho_{XB}=\sum_{x=0}^1 \hat p_{X}(x) \ket{x}\bra{x} \otimes \hat \rho_x^B \notag
\end{align}
such that $\hat \rho_X=\rho_X$ and $\hat \rho_B=\rho_B$. Then, we claim that $\hat \rho_{XB}=\rho_{XB}$. To see this,  note that  $\hat \rho_X=\rho_X$ implies that   $\hat p_{X}(0)=\hat p_{X}(1)=0.5$. Also, since $\hat \rho_B=\rho_B$ leads to $0.5 (\hat \rho_0+ \hat \rho_1)=\ket{\psi}\bra{\psi}$, the extremality of the pure state $\ket{\psi}\bra{\psi}$ in the convex set of density operators implies directly that
 $\hat \rho_0 =\hat \rho_1=\ket{\psi}\bra{\psi}$. Hence 
\begin{align}
    \theta^{\star}_{\mathsf{SL}}(\rho_{XB},\tilde \rho_{XB})&:= \min_{\substack{\hat \rho_{XB}:\\\hat \rho_{X}=\rho_X,~  \hat \rho_{B}=\rho_B} } \qrel{\hat \rho_{XB}}{\tilde \rho_{XB}} =  \qrel{\rho_{XB}}{\tilde \rho_{XB}}=\qrel{p_X}{\tilde p_X}+ \frac{1}{2} \big(\qrel{\psi}{\tilde \rho_0}+ \qrel{\psi}{\tilde \rho_1}\big). \notag
\end{align}

\subsection{Proof of Corollary \ref{Cor:upperbndsteinexp} }
The claim follows from \eqref{eq:zerorateweakconv} and 
\begin{align}
  &\theta^{\star}_{\mathsf{BSEP-ZR}}(\dutchcal{i}(p),\Phi^\perp) \leq \sup_{n  \in \NN}\sup_{\cM_n \in  \mathsf{SEP_n}}  \frac{\qrel{\cM_n\big(\dutchcal{i}(p)^{\otimes n}\big)}{\cM_n\big(\Phi^{\perp \otimes n}\big)}}{n}  \leq \log (pd+1), \notag \\
  & \theta^{\star}_{\mathsf{BSEP-ZR}}(\dutchcal{w}(p),\Theta) \leq \sup_{n  \in \NN}\sup_{\cM_n \in  \mathsf{SEP_n}}  \frac{\qrel{\cM_n\big(\dutchcal{w}(p)^{\otimes n}\big)}{\cM_n\big(\Theta^{\otimes n}\big)}}{n}  \leq \log \left(\frac{d+1-2p}{d-1}\right), \notag
\end{align}
where $\mathsf{SEP_n}$ denotes the set of all separable POVMs on $A^n:B^n$, and the final equalities in the equations  above follows from   \cite[Proposition 19]{RSB-IT-2025} and \cite[Proposition 25]{RSB-IT-2025}, respectively.

\subsection{Proof of Lemmas}
\subsubsection{Proof of Lemma \ref{Lem:qustwithclmarg}}\label{Sec:Lem:qustwithclmarg-proof}
For $p_X(x)>0$ and $p_Y(y)>0$, define
\[
\rho_A^x:=\frac{P_x^A\rho_AP_x^A}{p_X(x)},\qquad
\rho_B^y:=\frac{P_y^B\rho_BP_y^B}{p_Y(y)}.
\]
Analogous terms with $p_X(x)=0$ or $p_Y(y)=0$ may be omitted, since the marginal constraints force the corresponding $\hat p_{XY}(x,y)$ to vanish. Then the desired density operator $\hat\rho_{AB}$ is given as
\[
\hat\rho_{AB}:=\sum_{x,y}\hat p_{XY}(x,y)\rho_A^x\otimes\rho_B^y,
\]
where $\rho_A^x$ and $\rho_B^y$ are arbitrary density operators when $p_X(x)=0$ or $p_Y(y)=0$, respectively. 
Note that
\[
\hat\rho_A=\sum_xp_X(x)\rho_A^x=\sum_xP_x^A\rho_AP_x^A=\rho_A,
\]
where the last equality follows from $[P_x^A,\rho_A]=0$; similarly $\hat\rho_B=\rho_B$. Since $\rho_A^x$ and $\rho_B^y$ are supported on $P_x^A$ and $P_y^B$, respectively,
\[
\tr{(P_{x'}^A\otimes P_{y'}^B)\hat\rho_{AB}}=\hat p_{XY}(x',y').
\]
Observe that for rank-one marginal eigenbasis PVMs, $\rho_A^x=P_x^A$ and $\rho_B^y=P_y^B$.

\subsubsection{Proof of Lemma \ref{Lem:strong-converse-revhyp}} \label{Sec:Lem:strong-converse-revhyp-proof}
The proof relies on a reverse hypercontractivity result for the generalized depolarizing semigroup proved in \cite{Beigi-2020}. To introduce it, let $\cQ_{t,\rho}$  denote the following generalized depolarizing semigroup with invariant state $\rho$, i.e., for all bounded linear operators $L $,
\begin{align}
    \cQ_{t,\rho}(L) &\coloneqq e^{-t} L +  (1-e^{-t})\tr{L \rho} I. \notag
\end{align}
Let $(\cQ_{t,\rho}^n)_{t \geq 0}$ denote the  QMS obtained via tensor products, i.e., $  \cQ_{t,\rho}^n\coloneqq \bigotimes_{i=1}^n \cQ_{t,\rho}$.   
The following result is a special case of the reverse hypercontractivity for QMSs shown in \cite{Beigi-2020}, which will be sufficient for our purposes. 
\begin{theorem}[Reverse hypercontractivity for tensor products of depolarizing semigroup{\cite[Corollary 20]{Beigi-2020}}]\label{Thm:revhypcont}
  Consider the QMS 
$(\cQ_{t,\rho}^n)_{t \geq 0}$ with $\rho>0$ on the underlying Hilbert space. For any $p \leq q <1$ and $t \geq \log\left(\frac{p-1}{q-1}\right)$, the following holds:
\begin{align}
\norm{\cQ_{t,\rho}^n(M)}_{p,\rho^{\otimes n}} \geq \norm{M}_{q,\rho^{\otimes n}}, \quad \forall~ M>0, \notag
\end{align}
where for $r \in \RR \setminus \{0\}$,  $\norm{X}_{r,\sigma}\coloneqq \left(\tr{\abs{\sigma^{\frac{1}{2r}} X\sigma^{\frac{1}{2r}}}^r}\right)^{\frac{1}{r}}$ is the non-commutative weighted $L^r$ norm (pseudo-norm for $r<1$ since triangle inequality is not satisfied).
\end{theorem}
Continuing with the proof of  Lemma \ref{Lem:strong-converse-revhyp}, we may assume without loss of generality that  $V_n=B^n$ and $\cG_n^{B^n \to V^n}=\cI^{B^n \to B^n}$. If $\rho_A\not\ll\tilde\rho_A$ or $\rho_B\not\ll\tilde\rho_B$, then the right-hand side of \eqref{eq:type2err-strongconv-hypcont} is $+\infty$ by definition, and there is nothing to prove. Hence, in the nontrivial case we may assume $\rho_A\ll\tilde\rho_A$ and $\rho_B\ll\tilde\rho_B$.

First, assume that $\operatorname{supp}(\rho_A)=\operatorname{supp}(\tilde \rho_A)$ and $\operatorname{supp}(\rho_B)=\operatorname{supp}(\tilde \rho_B)$. 
Let  $\cF_n^{A^n \to W_n}$ be an arbitrary encoder (CPTP map),  
\begin{align}
 \sigma_{W_nB^n}:= \big(\cF_n^{A^n \to W_n} \otimes \cI^{B^n \to B^n}\big)(\rho_{AB}^{\otimes n}) \quad \mbox{and} \quad \tilde \sigma_{W_nB^n}=\tilde \sigma_{W_n}\otimes \tilde \rho_{B}^{\otimes n}:=\big(\cF_n^{A^n \to W_n} \otimes \cI^{B^n \to B^n}\big)\big(\tilde \rho_{A}^{\otimes n} \otimes \tilde \rho_{B}^{\otimes n}\big). \notag   
\end{align}
 Let $\cT_n=\{M_{W_nB^n},I_{W_nB^n}-M_{W_nB^n}\}$ be any  binary outcome POVM. Consider the  spectral decomposition: $\tilde \sigma_{W_n}=\sum_{w_n \in \cW_n} \lambda_{w_n} P_{w_n}$, where $\cP_{W_n}:=\{P_{w_n}\}_{w_n \in \cW_n}$ is a set of orthogonal rank-one projections such that $\sum_{w_n \in \cW_n}P_{w_n}=I_{W_n} $. Let  $\Pi_{W_n}(\cdot)$ denote the pinching map\footnote{Note that here we perform pinching with respect to rank-one orthogonal projections which is slightly different from the usual pinching operation, where the projections are formed by combining  eigenprojections corresponding to same  eigenvalues.}  using the projections in $\cP_{W_n}$, i.e.,
\begin{align}
  \Pi_{W_n}\big(\omega\big) =\sum_{w_n \in \cW_n}  P_{w_n} \omega P_{w_n}. \label{eq:pinchinzrmap}
\end{align}
A key idea in the proof is to use  pinching map to  construct a new test  $\bar \cT_n=\{\bar M_{W_nB^n}, I_{W_nB^n}-\bar M_{W_nB^n}\}$ satisfying two properties. Firstly, $\bar M_{W_nB^n}$ can be written as a separable sum, i.e., sum of tensor products of positive operators with the number of terms in the sum scaling at zero-rate with $n$. Secondly, the  pinching operation is such that the error probabilities achieved by $\bar \cT_n$ are not too different (in the exponent) from that achieved by $\cT_n$.   The separable decomposition of $\bar M_{W_nB^n}$  enables leveraging the reverse hypercontractivity result to relate the type II error probability with the type I error probability constraint $\epsilon$. The details are as follows. 

Setting
\begin{align}
&\bar M_{W_nB^n}:=\big(\Pi_{W_n} \otimes \cI^{B^n \to B^n} \big)\big(M_{W_nB^n}\big) =\sum_{w_n \in \cW_n}  \big(P_{w_n} \otimes I_{B^n}\big) M_{W_nB^n} \big(P_{w_n} \otimes I_{B^n}\big), \label{eq:modifiedpovm} 
\end{align}
 we have, using the pinching inequality of \cite{Hayashi_2002} (see also \cite{tomamichel2015quantum}), that
\begin{align}
    \bar M_{W_nB^n}    
    \geq \frac{M_{W_nB^n}}{|W_n|}.\label{eq:lowbndpovm}
\end{align}
Then, $\alpha_n(\cF_n,\cG_n,\cT_n) \leq \epsilon$ and \eqref{eq:lowbndpovm} implies
\begin{align}
    \tr{\sigma_{W_nB^n} \bar M_{W_nB^n} } \geq \frac{\tr{\sigma_{W_nB^n} M_{W_nB^n} }}{|W_n|} \geq \frac{1-\epsilon}{|W_n|}.\label{eq:lwrbndtr}
\end{align}
Next, observe that $ \bar M_{W_nB^n}$ can be written as
\begin{align}
  \bar M_{W_nB^n}=\sum_{w_n \in \cW_n} P_{w_n}^{W_n} \otimes M_{w_n}^{B^n},  
\end{align}
for some $\{M_{w_n}^{B^n}\}_{w_n \in \cW_n}$ such that $ M_{w_n}^{B^n} \geq 0$ for all $w_n \in \cW_n$. Hence 
\begin{align}
    \tr{\sigma_{W_nB^n} \bar M_{W_nB^n} } &=\tr{\big(\cF_n^{A^n \to W_n} \otimes \cI^{B^n \to B^n}\big)\big(\rho_{AB}^{\otimes n}\big)\bar M_{W_nB^n}} \notag \\
    &=\tr{\rho_{AB}^{\otimes n} \big(\cF_n^{\dag W_n \to A^n} \otimes \cI^{B^n \to B^n}\big)\big(\bar M_{W_nB^n}\big)} \notag \\
   & = \sum_{w_n \in \cW_n} \tr{  \left(\cF_n^{\dag W_n \to A^n}\big(P_{w_n}^{W_n}\big) \otimes M_{w_n}^{B^n}\right)\rho_{AB}^{\otimes n}} \notag \\
   &= \sum_{w_n \in \cW_n} \tr{  \big(M_{w_n}^{A^n} \otimes M_{w_n}^{B^n}\big)\rho_{AB}^{\otimes n}}, \label{eq:t2errprobpovm}
\end{align}
where $\cF_n^{\dag W_n \to A^n}$ is the adjoint map of $\cF_n^{A^n \to W_n}$ and $M_{w_n}^{A^n}:=\cF_n^{\dag W_n \to A^n}\big(P_{w_n}^{W_n}\big)$. Note that $0 \leq  M_{w_n}^{A^n} \leq I_{A^n}$ and $\sum_{w_n \in \cW_n} M_{w_n}^{A^n}=I_{A^n}$ since  $\cF_n^{\dag W_n \to A^n}$ is a completely positive unital map, being the adjoint of a CPTP map (see, e.g., \cite{tomamichel2015quantum}). 

From  \eqref{eq:lwrbndtr} and \eqref{eq:t2errprobpovm}, we obtain
\begin{align}
  \sum_{w_n \in \cW_n} \tr{  \big(M_{w_n}^{A^n} \otimes M_{w_n}^{B^n}\big)\rho_{AB}^{\otimes n}} \geq  \frac{1-\epsilon}{|W_n|}. \notag
\end{align} 
 Hence, there  exists some $w^{\star} \in \cW_n$  such that 
\begin{align}
    \tr{\rho_{AB}^{\otimes n} \big( M_{w^{\star}}^{A^n} \otimes M_{w^{\star}}^{B^n}  \big)}\geq \frac{1-\epsilon}{|W_n|^2}. \notag
\end{align}
It follows that
\begin{align}
\frac{1-\epsilon}{|W_n|^2}  &\leq   \tr{\rho_{AB}^{\otimes n} \big( M_{w^{\star}}^{A^n} \otimes M_{w^{\star}}^{B^n}  \big)} 
\leq  \tr{\rho_{AB}^{\otimes n} \big(M_{w^{\star}}^{A^n} \otimes I_{B^n}  \big)} = \tr{\rho_{A}^{\otimes n} M_{w^{\star}}^{A^n}}, \notag 
\end{align}
where the second inequality follows since $M_{w^{\star}}^{A^n} \otimes M_{w^{\star}}^{B^n} \leq  M_{w^{\star}}^{A^n} \otimes  I_{B^n}$ and  $\tr{AB} \leq \tr{AC}$ for $A \geq 0$ and $B \leq C$. Since a similar inequality also holds with $A^n$ replaced by the $B^n$ system, we have 
\begin{align}
    \tr{\rho_{A}^{\otimes n} M_{w^{\star}}^{A^n}} \wedge \tr{\rho_{B}^{\otimes n} M_{w^{\star}}^{B^n}} \geq \frac{1-\epsilon}{|W_n|^2}.\label{eq:lwrbndtrace}
\end{align}
Next, consider  the super-operator
\begin{align}
\tilde{\cQ}_{t,\rho,\tilde \rho}^n\coloneqq \bigotimes_{i=1}^n \tilde{\cQ}_{t,\rho,\tilde \rho}, \quad \mbox{with} \quad   \tilde{\cQ}_{t,\rho,\tilde \rho}(L) &\coloneqq e^{-t} L + \zeta(\rho,\tilde \rho) (1-e^{-t})\tr{L \tilde \rho} I,\quad \mbox{and} \quad  \zeta(\rho,\tilde \rho)\coloneqq \norm{\tilde{\rho}^{-\frac 12} \rho \tilde{\rho}^{-\frac 12}}_{\infty}. \notag  
\end{align}
 Then, we have
\begin{align}
\tr{\tilde \rho_B^{\otimes n}\tilde{\cQ}_{t,\rho_B,\tilde \rho_B}^n(M_{w^{\star}}^{B^n})}  \stackrel{(a)}{\geq} \tr{\tilde \rho_B^{\otimes n}\cQ_{t,\rho_B}^n(M_{w^{\star}}^{B^n})} \stackrel{(b)}{\geq} e^{-n\qrel{\rho_{B}}{\tilde \rho_B}} \left(\tr{\rho_{B}^{\otimes n}M_{w^{\star}}^{B^n}}\right)^{1+\frac{1}{t}}, \notag
\end{align}
where
\begin{enumerate}[label=(\alph*)]
    \item is since the super-operator $\tilde{\cQ}_{t,\rho,\tilde \rho}^n-\cQ_{t,\rho}^n$ is a positive map for all $t \geq 0$, i.e., $\tilde{\cQ}_{t,\rho,\tilde \rho}^n(L) \geq \cQ_{t,\rho}^n(L)$ for all $L \geq 0$ (see e.g. \cite{Beigi-2020,Cheng-2021});
  \item follows similar to \cite[Equation (50)]{Beigi-2020} utilizing Theorem \ref{Thm:revhypcont} and reverse H\"{o}lder inequality.  
\end{enumerate}
On the other hand, similar to \cite[Equation (51)]{Beigi-2020}, we have 
\begin{align}
    \tr{\tilde \rho_B^{\otimes n}\tilde{\cQ}_{t,\rho_B,\tilde \rho_B}^n(M_{w^{\star}}^{B^n})} \leq e^{n\big(\zeta(\rho_B,\tilde \rho_B)-1\big)t} \tr{\tilde \rho_B^{\otimes n}\tilde M_{w^{\star}}^{B^n}}. \notag
\end{align}
Combining the last two equations and \eqref{eq:lwrbndtrace} yields
\begin{subequations}\label{eq:lwrbndtrsyst}
\begin{align}
e^{n\big(\zeta(\rho_B,\tilde \rho_B)-1\big)t}\tr{\tilde \rho_B^{\otimes n}\tilde M_{w^{\star}}^{B^n}} \geq e^{-n\qrel{\rho_B}{\tilde \rho_B}} \left(\tr{\rho_{B}^{\otimes n}M_{w^{\star}}^{B^n}}\right)^{1+\frac{1}{t}} \geq e^{-n\qrel{\rho_B}{\tilde \rho_B}} \left(\frac{1-\epsilon}{|W_n|^2} \right)^{1+\frac{1}{t}}, \label{eq:lwrbndtrsystb}
\end{align}
and similarly
\begin{align}
 e^{n\big(\zeta(\rho_A,\tilde \rho_A)-1\big)t}\tr{\tilde \rho_A^{\otimes n}\tilde M_{w^{\star}}^{A^n}} \geq e^{-n\qrel{\rho_A}{\tilde \rho_A}} \left(\frac{1-\epsilon}{|W_n|^2} \right)^{1+\frac{1}{t}}. \label{eq:lwrbndtrsysta}  
\end{align}
\end{subequations}
Hence
\begin{align}
\beta_n(\cF_n,\cG_n,\cT_n)&\coloneqq \tr{\big(\tilde \sigma_{W_n}\otimes \tilde \rho_{B}^{\otimes n}\big)M_{W_nB^n}} \notag \\
    &\stackrel{(a)}{=}\tr{\big(\Pi_{W_n}\big(\tilde \sigma_{W_n}\big)\otimes \tilde \rho_{B}^{\otimes n}\big)M_{W_nB^n}}\notag \\
    &\stackrel{(b)}{=}\tr{\big(\tilde \sigma_{W_n}\otimes \tilde \rho_{B}^{\otimes n}\big)\bar{M}_{W_nB^n}} \notag \\
    &=\sum_{w_n \in \cW_n}\tr{\big(\cF_n(\tilde \rho_{A}^{\otimes n})\otimes \tilde \rho_{B}^{\otimes n}\big)\big(P_{w_n}^{W_n} \otimes M_{w_n}^{B^n}\big)} \notag \\
    & \stackrel{(c)}{=}\sum_{w_n \in \cW_n}\tr{\big(\tilde \rho_{A}^{\otimes n}\otimes \tilde \rho_{B}^{\otimes n}\big)\big(M_{w_n}^{A^n} \otimes M_{w_n}^{B^n}\big)} \notag \\
    &\stackrel{(d)}{\geq}\tr{\big(\tilde \rho_{A}^{\otimes n}\otimes \tilde \rho_{B}^{\otimes n}\big)\big(M_{w^{\star}}^{A^n} \otimes M_{w^{\star}}^{B^n}\big)} \notag \\
    &=\tr{M_{w^{\star}}^{B^n} \tilde \rho_{B}^{\otimes n}} \tr{M_{w^{\star}}^{A^n} \tilde \rho_{A}^{\otimes n}} \notag \\
    & \stackrel{(e)}{\geq} e^{-n\big(\qrel{\rho_A}{\tilde \rho_A}+\qrel{\rho_B}{\tilde \rho_B}+(\zeta(\rho_A,\tilde \rho_A)+\zeta(\rho_B,\tilde \rho_B)-2)t\big)} \left(\frac{1-\epsilon}{|W_n|^2} \right)^{2\left(1+\frac{1}{t}\right)}, \label{eq:type2errconv} 
\end{align}
where 
\begin{enumerate}[label=(\alph*)]
    \item is  because $\Pi_{W_n}\big(\tilde \sigma_{W_n}\big)=\tilde \sigma_{W_n}$ (see \eqref{eq:pinchinzrmap} for definition of $\Pi_{W_n}$);
    \item is using cyclicity of trace and  $\bar M_{W_nB^n}:=\big(\Pi_{W_n} \otimes \cI^{B^n \to B^n} \big)\big(M_{W_nB^n}\big)$;
    \item is again  via cyclicity of trace and  $M_{w_n}^{A^n}\coloneqq\cF_n^{\dag}\big(P_{w_n}^{W_n}\big)$;
    \item is due to $\tr{\big(\tilde \rho_{A}^{\otimes n}\otimes \tilde \rho_{B}^{\otimes n}\big)\big(M_{w_n}^{A^n} \otimes M_{w_n}^{B^n}\big)} \geq 0$ for all $w_n \in \cW_n$;
    \item is by applying \eqref{eq:lwrbndtrsyst}.
\end{enumerate}
Taking logarithms in \eqref{eq:type2errconv}, using $\zeta(\rho_A,\tilde\rho_A)+\zeta(\rho_B,\tilde\rho_B)-2\leq \zeta(\rho_A,\tilde\rho_A)+\zeta(\rho_B,\tilde\rho_B)$, and choosing the admissible value $t>0$
\begin{align}
t=\big(\xi(|W_n|,\epsilon)\big)^{\frac 12}\big(\zeta(\rho_A,\tilde \rho_A)+\zeta(\rho_B,\tilde \rho_B)\big)^{-\frac 12}n^{-\frac 12}, \notag 
\end{align}
gives the desired claim in \eqref{eq:type2err-strongconv-hypcont}.

We next consider the case when a null marginal $\operatorname{supp}(\rho_A)\subsetneq \operatorname{supp}(\tilde \rho_A)$ or $\operatorname{supp}(\rho_B)\subsetneq \operatorname{supp}(\tilde \rho_B)$.  For a pair $\rho\ll\tilde\rho$, restrict attention to $\operatorname{supp}(\tilde\rho)$ and let $\pi_{\tilde\rho}$ be the maximally mixed state on this support. For $\eta\in(0,1)$ define
\[
\rho^{(\eta)}:=(1-\eta)\rho+\eta\,\pi_{\tilde\rho}.
\]
Then $\rho^{(\eta)}>0$ on $\operatorname{supp}(\tilde\rho)$, so Theorem~\ref{Thm:revhypcont} applies with invariant state $\rho^{(\eta)}$. Importantly, this is only an auxiliary regularization in the reverse-hypercontractive estimate: the encoder, the test, the operators $M_{w_n}^{A^n},M_{w_n}^{B^n}$, and the index $w^\star$ selected below are those constructed from the original joint null state $\rho_{AB}^{\otimes n}$ and are kept fixed. Since $\rho^{(\eta)}\ge(1-\eta)\rho$, for every positive operator $L$,
\[
\tr{(\rho^{(\eta)})^{\otimes n}L}\ge (1-\eta)^n\tr{\rho^{\otimes n}L}.
\]
Using this and continuity of relevant quantities,
\[
D(\rho^{(\eta)}\|\tilde\rho)\to D(\rho\|\tilde\rho),\qquad
\zeta(\rho^{(\eta)},\tilde\rho)\to\zeta(\rho,\tilde\rho),
\]   we may apply reverse-hypercontractivity based arguments  first with $\rho^{(\eta)}$ that yields factors that converge to the respective unregularized quantities as $\eta\downarrow 0$ at fixed $n$. 
Hence,  \eqref{eq:type2err-strongconv-hypcont} can be shown to hold for all $\rho_A \ll \tilde \rho_A$ and $\rho_B \ll \tilde \rho_B$, thus completing the proof. 
\subsubsection{Proof of Lemma \ref{Lem:strong-converse}} \label{Sec:Lem:strong-converse-proof}
The proof of  Lemma \ref{Lem:strong-converse} will rely on the CPME-BL as stated below, which is a slight generalization of the classical blowing-up lemma  to 
quantum states $\rho_{AB}$ and $\sigma_{AB}$ with a product  eigenbasis of $\rho_A \otimes \rho_B$ commuting with $\sigma_{AB}$. 
 The classical blowing-up lemma    is a statement about concentration of measure at a high-level, and the CPME-BL also expresses a similar phenomenon at an operator level. It plays an analogous role as its classical counterpart in proving strong converses for distributed hypothesis testing problems  (see, e.g., \cite{Ahlswede-Csiszar,Shalaby-pap,SG_isit19,SD_ISIT2020,Sadaf-Tan-2021}). 
\begin{lemma}[Commuting product marginal eigenbasis blowing-up lemma (CPME-BL)]\label{lem:quant-blowup-bipartite}
    Suppose  $0< \epsilon_n \leq 1$,  $0 \leq M_n^{A^n} \otimes M_n^{B^n} \leq I_{A^nB^n}$, and $\rho_{AB} \in  \cS_{d_Ad_B}$ are such that $\tr{\rho_{A}^{\otimes n} M_n^{A^n}} \wedge \tr{\rho_{B}^{\otimes n} M_n^{B^n}} \geq \epsilon_n$ for all $n \in \NN$. Let $\sigma_{AB} \in \cS_{d_Ad_B}$ satisfy $\rho_A\otimes\rho_B\ll\sigma_{AB}$.  Moreover, suppose that there exist orthonormal eigenbasis $\{\ket{x}\}_{x\in\cX}$ of $\rho_A$ and $\{\ket{\bar x}\}_{\bar x\in\bar\cX}$ of $\rho_B$ such that $[\ket{x\bar x}\!\bra{x\bar x},\sigma_{AB}]=0$ for all $x\in\cX_+$ and $\bar x\in\bar\cX_+$. Then, for every non-negative sequence $(r_n)_{n \in \NN}$ there exists a projection $0 \leq  P_n^{+A^n} \otimes P_n^{+B^n} \leq I_{A^nB^n}$ such that 
\begin{subequations}
\begin{align}
    \tr{\rho_A^{\otimes n} P_n^{+A^n}} \wedge  \tr{\rho_B^{\otimes n} P_n^{+B^n}} &\geq 1-e^{-2r_n^2}, \label{eq:blowingup1}\\
    \tr{\sigma_{AB}^{\otimes n} \big( P_n^{+A^n} \otimes P_n^{+B^n}\big)} & \leq \gamma_n^2(\epsilon_n,r_n,\sigma_{AB}) \tr{\big(M_n^{A^n} \otimes M_n^{B^n} \big)\sigma_{AB}^{\otimes n}}, \label{eq:blowingup2}
\end{align}
$\text{where}$ 
\begin{align}
\gamma_n(\epsilon_n,r_n,\sigma_{AB})&:=\frac{2(d_A \vee d_B)^{ l_n(\epsilon_n,r_n)}\sum_{l=0}^{ l_n(\epsilon_n,r_n)}\binom{n}{l}}{\epsilon_n\big(\bar{\mu}_{\min}(\sigma_{AB})\big)^{ l_n(\epsilon_n,r_n)}},  \label{eq:blowupfactor} \\
l_n(\epsilon_n,r_n)&:=\left\lceil \sqrt{n}\left(\sqrt{-0.5 \log(0.5 \epsilon_n )}+r_n\right)\right\rceil, \label{eq:blowup-neighb} \\
\bar{\mu}_{\min}(\sigma_{AB})&:=\min_{x,\bar x \in \cX_+ \times \bar{\cX}_+ } \bra{x \bar x}\sigma_{AB}\ket{x \bar x}, \label{eq:mineigvalsup}
\end{align}
\end{subequations}
and $\cX_+$ (resp. $\bar\cX_+$) indexes the eigenvectors of $\rho_A$ (resp. $\rho_B$) with positive eigenvalues. 
\end{lemma}
The proof of Lemma \ref{lem:quant-blowup-bipartite} will be given in Appendix \ref{Sec:lem:quant-blowup-proof}. The key idea therein relies on identifying (based on $M_n^{A^n} \otimes M_n^{B^n}$) an appropriate subspace having sufficient overlap with the product marginals $\rho_{A}^{\otimes n} \otimes \rho_{B}^{\otimes n}$ and then leveraging the classical blowing-up lemma to construct projections $P_n^{+A^n}$ and $P_n^{+B^n}$ such that the two marginal overlaps are each at least $1-e^{-2r_n^2}$.  Consequently, for every $\hat\rho_{AB}^n\in\cD_n(\rho_{AB})$, the joint overlap satisfies
\[
\tr{\hat\rho_{AB}^n\big(P_n^{+A^n}\otimes P_n^{+B^n}\big)}
\ge 1-2e^{-2r_n^2},
\]
by the elementary union-bound/operator inequality used below.  At the same time, $\tr{\sigma_{AB}^{\otimes n} \big(P_n^{+A^n} \otimes P_n^{+B^n}\big)}$ differs negligibly (in the exponent) from $\tr{\sigma_{AB}^{\otimes n}\big(M_n^{A^n} \otimes M_n^{B^n}\big)}$. Based on the proof, we note that by construction, $\operatorname{supp}(P_n^{+A^n} \otimes P_n^{+B^n}) \subset \operatorname{supp}(\rho_A)^{\otimes n}\otimes\operatorname{supp}(\rho_B)^{\otimes n}$.

\medskip

Continuing with the proof of \eqref{eq:type2err-strongconv}, following similar steps as leading to \eqref{eq:t1errcl}, we have that $\alpha_n(\cF_n,\cG_n,\cT_n) \leq \epsilon$ implies 
\begin{align}
   \sum_{x \in \cX_n} \tr{\rho_{AB}^{\otimes n} \big( M_x^{A^n} \otimes M_x^{B^n}  \big)}\geq  1-\epsilon. \notag
\end{align} 
 Hence, there  exists some $x^{\star} \in \cX_n$  such that 
\begin{align}
    \tr{\rho_{AB}^{\otimes n} \big( M_{x^{\star}}^{A^n} \otimes M_{x^{\star}}^{B^n}  \big)}\geq \frac{1-\epsilon}{|X_n|}. \notag
\end{align}
Since $0 \leq M_{x^{\star}}^{A^n} \leq I_{A^n}$ and $0 \leq M_{x^{\star}}^{B^n} \leq I_{B^n}$, we have
\begin{align}
    \min\left\{\tr{\rho_{A}^{\otimes n}  M_{x^{\star}}^{A^n}}, \tr{\rho_{B}^{\otimes n} M_{x^{\star}}^{B^n}} \right\}\geq \frac{1-\epsilon}{|X_n|}. \label{eq:marginallowerbnd}
\end{align}
Next, we will use the CPME-BL to blow-up $M_{x^{\star}}^{A^n}$ and $M_{x^{\star}}^{B^n}$.  This lemma  implies that there exists $0 \leq P_n^{+A^n} \leq I_{A^n}$ and $0 \leq P_n^{+B^n}\leq I_{B^n}$   such that
\begin{align}
 & \min\left\{\tr{\rho_{A}^{\otimes n}  P_n^{+A^n}}, \tr{ \rho_{B}^{\otimes n} P_n^{+B^n}} \right\} \geq 1-e^{-2r_n^2}, \label{eq:err1probdeczero} \\
  \mbox{and } & \tr{\tilde \rho_{AB}^{\otimes n} \big( P_n^{+A^n} \otimes P_n^{+B^n}\big)}  \leq \gamma_n^2(\epsilon_n,r_n,\tilde \rho_{AB})\tr{\big(M_{x^{\star}}^{A^n} \otimes M_{x^{\star}}^{B^n} \big)\tilde \rho_{AB}^{\otimes n}} \leq \gamma_n^2(\epsilon_n,r_n,\tilde \rho_{AB}) \mspace{2 mu}\beta_n(\cF_n,\cG_n,\cT_n), \notag
\end{align}
where $\gamma_n(\epsilon_n,r_n,\tilde \rho_{AB})$ is as defined in \eqref{eq:blowupfactor}  with $\tilde \rho_{AB}$ in place of $\sigma_{AB}$ and 
\begin{align}
l_n( \epsilon_n,r_n)&:=\sqrt{n}\left(\sqrt{-0.5 \log\big(0.5 \epsilon_n \big)}+r_n\right), \quad \mbox{ with } \quad  \epsilon_n:=\frac{1-\epsilon}{|X_n|}. \label{eq:Hammneighsize} 
\end{align}

   Since $\hat \rho^n_{A}=\rho_A^{\otimes n}$ and $\hat \rho^n_{B}=\rho_B^{\otimes n}$ for any $\hat \rho^n_{AB} \in \cD_n(\rho_{AB})$, we have
\begin{align}
   \min_{\hat \rho^n_{AB} \in \cD_n(\rho_{AB})} \min\left\{\tr{\hat \rho_{A}^n  P_n^{+A^n}}, \tr{\hat \rho_{B}^n P_n^{+B^n}} \right\} \geq  1-e^{-2r_n^2}.   \notag
\end{align}

Next, note that for any commuting $M_1,M_2$ such that $0 \leq M_1,M_2 \leq I$, we have  $0 \leq (I-M_1)(I-M_2) \leq I$. Hence, for such $M_1,M_2$ and any density operator $\sigma $, $\tr{\sigma(I-M_1)(I-M_2)} \geq 0$, i.e.,
\begin{align}
   1-\tr{\sigma M_1}-\tr{\sigma M_2}+\tr{\sigma M_1M_2} \geq  0. \label{eq:probintersecevent}
\end{align}
Applying this with $M_1=P_n^{+A^n} \otimes I_{B^n}$ and $M_2=I_{A^n} \otimes P_n^{+B^n}$,  
we have for any $\hat \rho^n_{AB} \in \cD_n(\rho_{AB})$ that
\begin{align}
   \tr{\hat \rho^n_{AB}  (P_n^{+A^n} \otimes P_n^{+B^n})}&=\tr{\hat \rho^n_{AB}  (P_n^{+A^n} \otimes I_{B^n})(I_{A^n} \otimes P_n^{+B^n})} \notag \\
    & \geq \tr{\hat \rho^n_{AB}  (P_n^{+A^n} \otimes I_{B^n})}+\tr{\hat \rho^n_{AB}  ( I_{A^n} \otimes P_n^{+B^n})}-1 \notag \\
    &=\tr{\hat \rho_{A}^n  P_n^{+A^n}}+ \tr{\hat \rho_{B}^n  P_n^{+B^n}}-1 \notag \\
    & \geq 1-2 e^{-2r_n^2}.\label{eq: type1errprobminset}
\end{align}
Let $0 \leq \hat \beta_n(\tilde \rho_{AB}):=\tr{\tilde  \rho_{AB}^{\otimes n}  (P_n^{+A^n} \otimes P_n^{+B^n})} \leq 1$ and $0 \leq  \hat \alpha_n (\hat \rho^n_{AB}):=\tr{\hat \rho^n_{AB}  (I-P_n^{+A^n} \otimes P_n^{+B^n})}\leq 1$.    Consider the local POVM $\cM_n^+:=\cM_+^{A^n} \otimes \cM_+^{B^n}$, where $\cM_+^{A^n} :=\{P_n^{+A^n}, I-P_n^{+A^n} \}$ and $  \cM_+^{B^n}:= \{P_n^{+B^n}, I-P_n^{+B^n}\}$ are binary outcome POVMs. 
By log-sum inequality, we have
\begin{align}
    \qrel{\cM_n^+(\hat \rho^n_{AB})}{\cM_n^+(\tilde  \rho_{AB}^{\otimes n})} & \geq \tr{\hat \rho^n_{AB}  (P_n^{+A^n} \otimes P_n^{+B^n})} \log \left(\frac{\tr{\hat \rho^n_{AB}  (P_n^{+A^n} \otimes P_n^{+B^n})}}{\tr{\tilde \rho_{AB}^{\otimes n}  (P_n^{+A^n} \otimes P_n^{+B^n})}}\right) \notag \\
    &\quad +\tr{\hat \rho^n_{AB}  (I-P_n^{+A^n} \otimes P_n^{+B^n})} \log \left(\frac{\tr{\hat \rho^n_{AB}  (I-P_n^{+A^n} \otimes P_n^{+B^n})}}{\tr{\tilde \rho_{AB}^{\otimes n}  (I-P_n^{+A^n} \otimes P_n^{+B^n})}}\right) \notag \\
    &=-h_b\big(\hat \alpha_n(\hat \rho^n_{AB})\big)-\big(1-\hat \alpha_n(\hat \rho^n_{AB})\big) \log \big(\hat \beta_n(\tilde \rho_{AB})\big)-\hat \alpha_n (\hat \rho^n_{AB})\log\big(1-\hat \beta_n(\tilde \rho_{AB})\big) \notag \\
    & \geq -h_b\big(\hat \alpha_n(\hat \rho^n_{AB})\big)-\big(1-\hat \alpha_n(\hat \rho^n_{AB})\big) \log \big(\hat \beta_n(\tilde \rho_{AB})\big) \notag \\
        & \geq -h_b\big(\hat \alpha_n(\hat \rho^n_{AB})\big)-\big(1-\hat \alpha_n(\hat \rho^n_{AB})\big) \big(\log \beta_n(\cF_n,\cG_n,\cT_n)+2\log \gamma_n(\epsilon_n,r_n,\tilde \rho_{AB})\big), \notag
\end{align}
where $h_b$ is the binary entropy function. 
Hence, we have 
\begin{align}
    -\log \beta_n(\cF_n,\cG_n,\cT_n)& \leq \frac{\qrel{\cM_n^+(\hat \rho^n_{AB})}{\cM_n^+(\tilde  \rho_{AB}^{\otimes n})}+ h_b\big(\hat \alpha_n(\hat \rho^n_{AB})\big)}{\big(1-\hat \alpha_n(\hat \rho^n_{AB})\big)} +2 \log \gamma_n(\epsilon_n,r_n,\tilde \rho_{AB}).  \notag
\end{align}
From \eqref{eq: type1errprobminset}, it follows that
\begin{align}
    -\frac{1}{n}\log \beta_n(\cF_n,\cG_n,\cT_n)\leq \frac{\qrel{\cM_n^+(\hat \rho^n_{AB})}{\cM_n^+(\tilde  \rho_{AB}^{\otimes n})}}{n(1-2e^{-2r_n^2})}+\frac{2}{n}\log \gamma_n(\epsilon_n,r_n,\tilde \rho_{AB})+\frac{1}{n(1-2e^{-2r_n^2})}.  \notag
\end{align}
Since the above holds for all $\hat \rho^n_{AB} \in \cD_n(\rho_{AB})$, we obtain by taking infimum that
\begin{align}
 -\frac{1}{n}\log \beta_n(\cF_n,\cG_n,\cT_n) &\leq  \inf_{\hat \rho^n_{AB} \in \cD_n(\rho_{AB})}\frac{\qrel{\cM_n^+(\hat \rho^n_{AB})}{\cM_n^+(\tilde  \rho_{AB}^{\otimes n})}}{n(1-2e^{-2r_n^2})}+ \frac{2}{n}\log \gamma_n(\epsilon_n,r_n,\tilde \rho_{AB})+\frac{1}{n(1-2e^{-2r_n^2})}. \notag
\end{align}
Since $P_n^{+A^n}$ and $P_n^{+B^n}$ are sums of rank-one eigenprojectors of $\rho_A^{\otimes n}$ and $\rho_B^{\otimes n}$, respectively, the constructed binary local PVM belongs to $\mathsf{BPLO}^{\rho}_n$. Taking the supremum over $\cM_n^+\in\mathsf{BPLO}^{\rho}_n$ therefore gives
\begin{align}
   -\frac{1}{n}\log \beta_n(\cF_n,\cG_n,\cT_n) &\leq  \sup_{\cM_n \in \mathsf{BPLO}^{\rho}_n}\,\inf_{\hat \rho^n_{AB} \in \cD_n(\rho_{AB})}\frac{\qrel{\cM_n(\hat \rho^n_{AB})}{\cM_n(\tilde  \rho_{AB}^{\otimes n})}+1}{n(1-2e^{-2r_n^2})}+  \frac{2}{n}\log \gamma_n(\epsilon_n,r_n,\tilde \rho_{AB}).  \notag
\end{align}
Since the RHS is independent of $\cF_n,\cG_n,\cT_n$,  we obtain \eqref{eq:type2err-strongconv}.
\section{Concluding Remarks}\label{Sec:concrem}
We considered a quantum distributed hypothesis testing problem under zero-rate noiseless communication constraint and characterized the Stein's exponent in terms of an efficiently computable single-letter expression, when the state under the alternative is a product of its marginals. We also obtained multi-letter characterizations of the Stein's exponent in the general case when at least one of the parties communicates classically at zero-rate. Under a support condition and product eigenbasis commutativity condition, our exponent can be stated in terms of max-min optimization of regularized measured relative entropy, where the maximization is over null marginal compatible local measurements and minimization is over auxiliary states with the same marginals as the tensor-product state under the null hypothesis. 
We also showed via an example of classical-quantum states that the Stein's exponent does not coincide with the natural quantum analogue of the single-letter expression characterizing it classically.

Several open questions still remain in the context of zero-rate quantum distributed hypothesis testing. One such question is to characterize the Stein's exponent for general correlated CQ states that does not satisfy the commuting product marginal eigenbasis condition.   Of interest is also a fully quantum version of the blowing-up lemma. Another question pertains to the Stein's exponent with two-sided  quantum communication at zero-rate, i.e.,  when both Alice and Bob communicates quantum information to Charlie. Here, we solved a special case of this problem when the state under the alternative is the product of its  marginals; however, the general case remains open.    Another important problem is the trade-off between the (Hoeffding's) exponents  of both the type I and type II error probabilities  as well as the  Chernoff's exponent. Also relevant is the strong-converse exponent in zero-rate settings which quantifies how fast the success probability of one type decays to zero given that the exponent of the other error probability is above the Stein's exponent. An answer to these questions would provide a complete account of asymptotic performance  for quantum distributed hypothesis testing under zero-rate noiseless communication. 

Performance guarantees for distributed quantum hypothesis testing problems are useful  for benchmarking the fundamental limitations (or capabilities) of  quantum-assisted artificially intelligent systems since hypothesis testing is a primitive that underlies many inference tasks.   Zero-rate communication is  practically relevant in scenarios where communication is  expensive, e.g., due to a low energy budget. For instance, wireless sensor networks powered by batteries used for industrial automation, agriculture, smart cities, environmental monitoring or healthcare have  limited energy budget and typically monitor large amounts of data  between successive communication phases with the central controller. Potential applications of our study include benchmarking the performance of tasks in quantum sensing and metrology,  which  benefit from a spatially distributed sensor network architecture (see, e.g., \cite{Proctor-17, Zhan-23,Rivero-2021} and references therein). Another pertinent application is  covert inference,  where  zero-rate communication becomes essential to avoid detection by an adversary (see, e.g., \cite{BGPHTG-2015,Bash-2013, Wang-2016,Bloch-2016}).  

\section*{Acknowledgments}
MB acknowledges funding by the European Research Council (ERC Grant Agreement No. 948139). MB and SS acknowledges support from the Excellence Cluster - Matter and Light for Quantum Computing (ML4Q). HC is supported by NSTC 113-2119-M-001-009, No.~NSTC 113-2628-E-002-029, No.~NTU-113V1904-5, No.~NTU-CC-113L891605, and No.~NTU-113L900702.
\begin{appendices} 
\section{Supremum and Infimum in \eqref{eq:optimalmultlettexp} are achieved under support condition}\label{App:supinfach}
Let $P_n:=\operatorname{supp}\!\big(\rho_A^{\otimes n}\otimes\rho_B^{\otimes n}\big)$. 
For every $\hat\rho^n_{AB}\in\cD_n(\rho_{AB})$, the support of $\hat\rho^n_{AB}$ is contained in $P_n$.
Since $\rho_A\otimes\rho_B\ll\tilde\rho_{AB}$, we also have
$P_n\ll\tilde\rho_{AB}^{\otimes n}$. Hence
\[
c_n:=
\left\|
(\tilde\rho_{AB}^{\otimes n})^{-1/2}
P_n
(\tilde\rho_{AB}^{\otimes n})^{-1/2}
\right\|_\infty<\infty,
\qquad
P_n\le c_n\,\tilde\rho_{AB}^{\otimes n},
\]
where the inverse is taken on $\operatorname{supp}(\tilde\rho_{AB}^{\otimes n})$.
Moreover, because $\hat\rho^n_{AB}$ is a density operator supported on $P_n$,
$\hat\rho^n_{AB}\le P_n$, and therefore
\[
\hat\rho^n_{AB}\le c_n\,\tilde\rho_{AB}^{\otimes n}.
\]
After any measurement $\cM_n$, this yields the outcome-wise domination
$p_i\le c_n q_i$, where
$p_i$ and $q_i$ are the probabilities obtained by measuring
$\hat\rho^n_{AB}$ and $\tilde\rho_{AB}^{\otimes n}$, respectively.
Thus the classical relative entropy
$D(\cM_n(\hat\rho^n_{AB})\|\cM_n(\tilde\rho_{AB}^{\otimes n}))$
is finite and jointly continuous in $(\cM_n,\hat\rho^n_{AB})$ on the relevant compact sets. 

Define
\[
f(\cM_n,\hat\rho^n_{AB})
:=D\!\left(\cM_n(\hat\rho^n_{AB})\middle\|\cM_n(\tilde\rho_{AB}^{\otimes n})\right),
\qquad
g(\cM_n):=\min_{\hat\rho^n_{AB}\in\cD_n(\rho_{AB})}f(\cM_n,\hat\rho^n_{AB}).
\]
Moreover,  $\cD_n(\rho_{AB})$ is nonempty, compact, and independent of $\cM_n \in \mathsf{PLO}^{\rho}_n$.   The preceding domination argument implies that $f$ is jointly continuous on the relevant compact domain. Hence Berge's maximum theorem (equivalently, its minimum version applied to $-f$) implies directly that the value function $g$ is continuous. In particular, the inner minimum is attained for every $\cM_n$. Finally, $\mathsf{PLO}^{\rho}_n$ is compact (as a closed subset of the finite-dimensional compact set of rank-one local PVM measurement channels), so the outer supremum in \eqref{eq:optimalmultlettexp} is also attained.
\section{Proof of Lemma \ref{lem:quant-blowup-bipartite}}\label{Sec:lem:quant-blowup-proof}
Note that the support condition $\rho_A\otimes\rho_B\ll\sigma_{AB}$ together with $[\ket{x\bar x}\!\bra{x\bar x},\sigma_{AB}]=0$ on $\cX_+\times\bar\cX_+$  implies $\bra{x\bar x}\sigma_{AB}\ket{x\bar x}>0$ for every $x\in\cX_+$ and $\bar x\in\bar\cX_+$, and hence $\bar\mu_{\min}(\sigma_{AB})>0$. 
 Consider the spectral decompositions $\rho_A=\sum_{x_i \in \cX}\lambda_{x_i} \ket{x_i}\bra{x_i}$ and $\rho_B=\sum_{\bar x_i \in \bar {\cX}}\bar{\lambda}_{\bar x_i} \ket{\bar x_i}\bra{\bar  x_i}$. Then, we have  $\rho_A^{\otimes n}=\sum_{x^n \in \cX^n } \lambda_{x^n} \ket{x^n}\bra{x^n}$ and $\rho_B^{\otimes n}=\sum_{\bar x^n \in \bar {\cX}^n } \bar {\lambda}_{\bar  x^n} \ket{\bar  x^n}\bra{\bar 
 x^n}$. Set $\cX_+^n:=\{x^n \in \cX^n: \lambda_{x^n} >0\}$ and $\bar{\cX}_+^n:=\{\bar x^n \in \bar{\cX}^n: \lambda_{\bar x^n} >0\}$. Let $l_n:=l_n(\epsilon_n,r_n)$ be as defined in \eqref{eq:blowup-neighb}, and 
 \begin{align}
\cJ_n(M_n^{A^n},\epsilon_n)&:=\{x^n \in \cX_+^n: \bra{x^n}M_n^{A^n}\ket{x^n} 
 \geq 0.5\epsilon_n\}, \notag \\
\bar \cJ_n(M_n^{B^n},\epsilon_n)& :=\{\bar x^n \in \bar{\cX}_+^n: \bra{\bar x^n}M_n^{B^n}\ket{\bar x^n} 
 \geq 0.5\epsilon_n\}. \notag 
 \end{align}
 Since $ \tr{\rho_A^{\otimes n}M_n^{A^n}} \geq \epsilon_n$ by assumption, we have
\begin{align}
\epsilon_n \leq \tr{\rho_A^{\otimes n}M_n^{A^n}}&=\sum_{x^n \in \cJ_n(M_n^{A^n},\epsilon_n)} \lambda_{x^n} \bra{x^n}M_n^{A^n}\ket{x^n}+  \sum_{x^n \in \cX_+^n \setminus \cJ_n(M_n^{A^n},\epsilon_n)} \lambda_{x^n} \bra{x^n}M_n^{A^n}\ket{x^n} \notag \\
 &  \leq  \sum_{x^n \in \cJ_n(M_n^{A^n},\epsilon_n)} \lambda_{x^n} \bra{x^n}M_n^{A^n}\ket{x^n}+{0.5}\,\epsilon_n \sum_{x^n \in \cX_+^n \setminus \cJ_n(M_n^{A^n},\epsilon_n)} \lambda_{x^n} \notag \\
 & \leq \sum_{x^n \in \cJ_n(M_n^{A^n},\epsilon_n)} \lambda_{x^n} \bra{x^n}M_n^{A^n}\ket{x^n}+0.5 \epsilon_n \notag \\
 & \leq \sum_{x^n \in \cJ_n(M_n^{A^n},\epsilon_n)} \lambda_{x^n} +0.5 \epsilon_n, \notag
\end{align}
since $\bra{x^n}M_n^{A^n}\ket{x^n} \leq 1$ for all $x^n \in \cX^n$. 
Hence, with $ P_n^{A^n}=\sum_{x^n \in \cJ_n(M_n^{A^n},\epsilon_n)}\ket{x^n} \bra{x^n}$, we have
\begin{align}
\tr{\rho_A^{\otimes n} P_n^{A^n}}=\sum_{x^n \in \cJ_n(M_n^{A^n},\epsilon_n)} \lambda_{x^n} \geq 0.5\epsilon_n.\notag
\end{align}
Similarly, using $ \tr{\rho_B^{\otimes n}M_n^{B^n}} \geq \epsilon_n$, we have $\tr{\rho_B^{\otimes n} P_n^{B^n}}\geq 0.5\epsilon_n$ with $ P_n^{B^n}=\sum_{\bar x^n \in \cJ_n(M_n^{B^n},\epsilon_n)}\ket{\bar x^n} \bra{\bar x^n}$.

For  $r_n \geq 0$, let  $l_n:=l_n(\epsilon_n,r_n):=\big\lceil \sqrt{n}\big(\sqrt{-0.5 \log(0.5 \epsilon_n )}+r_n\big)\big\rceil$ be as defined in \eqref{eq:blowup-neighb}. We will use the classical blowing-up lemma \cite{Ahlswede-Gacs-Korner-1976,Marton-1986} to blow-up the projections $P_n^{A^n}$ and $P_n^{B^n}$ such that they contain most of the densities of $\rho_A^{\otimes n}$ and $\rho_B^{\otimes n}$, respectively. Denote the  Hamming $l_n$ neighborhoods of the $\cJ_n(M_n^{A^n},\epsilon_n)$ and $\cJ_n(M_n^{B^n},\epsilon_n)$ by $\cJ_n^{l_n+}(M_n^{A^n},\epsilon_n)$ and $\bar \cJ_n^{l_n+}(M_n^{B^n},\epsilon_n)$, respectively, where the Hamming neighborhoods are taken inside the support alphabets $\cX_+^n$ and $\bar\cX_+^n$. Thus, for $\tilde\cX_n\subseteq\cX_+^n$ its Hamming $l$-neighborhood is
$\{x^n\in\cX_+^n:\exists\,\hat x^n\in\tilde\cX_n\text{ such that }d_{\mathsf H}(x^n,\hat x^n)\le l\}$, and analogously on $B^n$.
Here $d_{\mathsf H}(x^n,\hat x^n):=\sum_{i=1}^n\ind_{x_i\neq\hat x_i}$ denotes the Hamming distance. 

Setting
\begin{align}
P_n^{+A^n}:=\sum_{x^n \in \cJ_n^{l_n+}(M_n^{A^n},\epsilon_n)} \ket{x^n} \bra{x^n} \qquad \mbox{ and } \qquad  
P_n^{+B^n}:=\sum_{\bar x^n \in \bar \cJ_n^{l_n+}(M_n^{B^n},\epsilon_n)} \ket{\bar x^n} \bra{\bar x^n}, \notag
 \end{align} 
 we obtain using the version of blowing-up lemma given in \cite[Lemma 3.6.2]{Raginsky-Sason-2013} that 
 \begin{subequations} \label{eq:blupprobone}
 \begin{align}
\tr{\rho_A^{\otimes n} P_n^{+A^n}}=\sum_{x^n \in \cJ_n^{l_n+}(M_n^{A^n},\epsilon_n)} \lambda_{x^n} \geq 1-e^{-2r_n^2}, \\
\tr{\rho_B^{\otimes n} P_n^{+B^n}}=\sum_{\bar x^n \in \cJ_n^{l_n+}(M_n^{B^n},\epsilon_n)} \lambda_{\bar x^n} \geq 1-e^{-2r_n^2}.
\end{align}
 \end{subequations} 
thus showing \eqref{eq:blowingup1}. Next, note that 
 for $x^n,\hat x^n, \bar x^n, \check x^n$ such that $d_{\mathsf{H}}( x^n,\hat x^n) \vee d_{\mathsf{H}}(\bar x^n, \check x^n)\leq l_n $, the Hamming distance between the pairs $(x^n, \bar x^n)$ and $(\hat x^n,\check x^n)$ is at most $2 l_n$. Hence,  we have
\begin{align}
    \prod_{i=1}^n \bra{\hat x_i \check x_i}\sigma_{AB}\ket{\hat x_i \check x_i} & \leq \prod_{i=1}^n \bra{x_i \bar x_i}\sigma_{AB}\ket{x_i\bar x_i} \left( \frac{\max_{x,\bar x \in \cX_+ \times \bar{\cX}_+ } \bra{x \bar x}\sigma_{AB}\ket{x \bar x}}{\min_{x,\bar x \in \cX_+ \times \bar{\cX}_+ } \bra{x \bar x}\sigma_{AB}\ket{x \bar x}}\right)^{2l_n} \notag \\
    & \leq \prod_{i=1}^n \bra{x_i \bar x_i}\sigma_{AB}\ket{x_i\bar x_i} \left( \frac{1}{\bar \mu_{\min}(\sigma_{AB})}\right)^{2l_n} .  \notag 
\end{align}
Moreover, there are at most $d_A^{l_n}\sum_{l=0}^{l_n}\binom nl$ and $d_B^{l_n}\sum_{l=0}^{l_n}\binom nl$  sequences in an  $l_n$ Hamming neighborhood of any sequence $x^n$ and $\bar x^n$ sequence, respectively.
Thus, we have 
\begin{align}
 & \tr{\sigma_{AB}^{\otimes n} \big(P_n^{+A^n} \otimes P_n^{+B^n}\big)} \notag \\
 &=\sum_{\hat x^n \in \cJ_n^{l_n+}(M_n^{A^n},\epsilon_n)} \sum_{\check x^n \in \bar \cJ_n^{l_n+}(M_n^{B^n},\epsilon_n)} \prod_{i=1}^n \bra{\hat x_i \check x_i }\sigma_{AB}\ket{\hat x_i \check x_i} \notag \\
   & \leq \left((d_A\vee d_B)^{l_n}\sum_{l=0}^{l_n}\binom nl\right)^2\left( \frac{1}{\bar \mu_{\min}(\sigma_{AB})}\right)^{2l_n}\sum_{x^n \in \cJ_n(M_n^{A^n},\epsilon_n)} \sum_{\bar x^n \in \bar \cJ_n(M_n^{B^n},\epsilon_n)} \prod_{i=1}^n \bra{x_i \bar x_i}\sigma_{AB}\ket{x_i \bar x_i} \notag\\
     & \leq \frac{4}{\epsilon_n^2}\left((d_A\vee d_B)^{l_n}\sum_{l=0}^{l_n}\binom nl\right)^2\left( \frac{1}{\bar \mu_{\min}(\sigma_{AB})}\right)^{2l_n}\sum_{x^n \in \cJ_n(M_n^{A^n},\epsilon_n)} \sum_{\bar x^n \in \bar \cJ_n(M_n^{B^n},\epsilon_n)} \bra{x^n \bar x^n}(M_n^{A^n} \otimes M_n^{B^n})\ket{x^n \bar x^n} \notag \\
     & \qquad \qquad \qquad \qquad \qquad \qquad  \qquad \qquad \qquad  \qquad \qquad \qquad  \qquad \qquad \qquad  \qquad \qquad \qquad \times \prod_{i=1}^n \bra{x_i \bar x_i}\sigma_{AB}\ket{x_i \bar x_i} \notag\\
  & = \gamma_n^2(\epsilon_n,r_n,\sigma_{AB})\sum_{x^n \in \cJ_n(M_n^{A^n},\epsilon_n)} \sum_{\bar x^n \in \bar \cJ_n(M_n^{B^n},\epsilon_n)} \bra{x^n \bar x^n}(M_n^{A^n} \otimes M_n^{B^n})\ket{x^n \bar x^n}\prod_{i=1}^n \bra{x_i \bar x_i}\sigma_{AB}\ket{x_i \bar x_i} \notag\\
  & = \gamma_n^2(\epsilon_n,r_n,\sigma_{AB})\sum_{x^n \in \cJ_n(M_n^{A^n},\epsilon_n)} \sum_{\bar x^n \in \bar \cJ_n(M_n^{B^n},\epsilon_n)} \tr{(M_n^{A^n} \otimes M_n^{B^n})\ket{x^n \bar x^n}\bra{x^n \bar x^n}\sigma_{AB}^{\otimes n}\ket{x^n \bar x^n} \bra{x^n \bar x^n}} \notag\\
   & \stackrel{(a)}{=} \gamma_n^2(\epsilon_n,r_n,\sigma_{AB})\sum_{x^n \in \cJ_n(M_n^{A^n},\epsilon_n)} \sum_{\bar x^n \in \bar \cJ_n(M_n^{B^n},\epsilon_n)} \tr{(M_n^{A^n} \otimes M_n^{B^n})\sigma_{AB}^{\otimes n}\ket{x^n \bar x^n} \bra{x^n \bar x^n}} \notag\\
   &=\gamma_n^2(\epsilon_n,r_n,\sigma_{AB}) \tr{(M_n^{A^n} \otimes M_n^{B^n})\sigma_{AB}^{\otimes n}\big(P_n^{+A^n} \otimes P_n^{+B^n}\big)} \notag\\
   & \stackrel{(b)}{\leq} \gamma_n^2(\epsilon_n,r_n,\sigma_{AB}) \tr{(M_n^{A^n} \otimes M_n^{B^n})\sigma_{AB}^{\otimes n}}, \notag
\end{align}
where $\gamma_n(\epsilon_n,r_n,\sigma_{AB})$ is as defined in  \eqref{eq:blowupfactor}. In the above, $(a)$ follows from the commuting product marginal eigenbasis  assumption: each one-copy projector $\ket{x\bar x}\!\bra{x\bar x}$ commutes with $\sigma_{AB}$, and therefore every tensor-product projector $\ket{x^n\bar x^n}\!\bra{x^n\bar x^n}$ commutes with $\sigma_{AB}^{\otimes n}$. Since $P_n^{+A^n}\otimes P_n^{+B^n}$ is a sum of these mutually orthogonal projectors, it also commutes with $\sigma_{AB}^{\otimes n}$. This implies that  $0\leq \sigma_{AB}^{\otimes n} \big(P_n^{+A^n}\otimes P_n^{+B^n} \big) \leq \sigma_{AB}^{\otimes n}$, from which the final inequality follows using $\tr{AB} \leq \tr{AC}$ for $A,B,C \geq 0$ and $B \leq C$. 
\end{appendices}

\bibliographystyle{IEEEtran}
\bibliography{ref,ref-quant,ref2}

\end{document}